\documentclass[usenatbib]{mn2e}
\usepackage{graphicx,times}
\usepackage{bm}
\usepackage{epsf}
\usepackage{amsmath}
\usepackage{amssymb}

\date{\today,~ $ $Revision: 0.9 $ $}

\def\la{\langle}
\def\ra{\rangle}
\def\n{\noindent}
\def\be{\begin{equation}}
\def\ee{\end{equation}}
\def\ben{\begin{eqnarray}}
\def\een{\end{eqnarray}}
\def\nn{\nonumber}
\def\oh{\hat\Omega}
\def\myC{{\cal C}}
\def\myB{{\cal B}}
\def\myf{\kappa}
\def\bk{{\bf k}}

\def\br{{\bf r}}

\def\myC{{\cal C}}
\def\myB{{\cal B}}
\def\myT{{\cal T}}

\def\ad{{_{\rm{lin}}}}

\def\bk{{\bf k}}

\def\2p{{(2\pi)^2}}

\def\be{\begin{equation}}
\def\ee{\end{equation}}

\def\beq{\begin{equation}}
\def\eeq{\end{equation}}

\def\ben{\begin{eqnarray}}
\def\een{\end{eqnarray}}

\def\oh{{\hat\Omega}}
\def\nn{{\nonumber}}

\newcommand{\beqa}{\begin{eqnarray}}
\newcommand{\eeqa}{\end{eqnarray}}

\begin{document}

\onecolumn
\title[Weak Lensing, non-Gaussianity and Minkowski Functionals]
{From Weak Lensing to non-Gaussianity via Minkowski Functionals}

\author[Munshi et al.]
{Dipak Munshi$^{1}$, Ludovic van Waerbeke$^{2}$, Joseph Smidt$^{3}$, Peter Coles$^{1}$ \\
$^{1}$School of Physics and Astronomy, Cardiff University, Queen's
Buildings, 5 The Parade, Cardiff, CF24 3AA, UK\\
$^{2}$Department of Physics \& Astronomy, University of British Columbia,
6224 Agricultural Road, Vancouver, B.C. V6T 1Z1, Canada\\
$^{3}$ Department of Physics and Astronomy, University of California, Irvine, CA 92697}
\maketitle

\begin{abstract}
We present a new harmonic-domain based approach for extracting
morphological information, in the form of Minkowski Functionals
(MFs), from weak lensing (WL) convergence maps. Using a perturbative
expansion of the MFs, which is expected to be valid for the range of
angular scales probed by most current weak-lensing surveys, we show
that the study of three generalized  skewness parameters is
equivalent to the study of the three MFs defined in two dimensions.
We then extend these skewness parameters to three associated
skew-spectra which carry more information about the convergence
bispectrum than their one-point counterparts. We discuss various
issues such as noise and incomplete sky coverage in the context of
estimation of these skew-spectra from realistic data. Our technique
provides an alternative to the pixel-space approaches typically used
in the estimation of MFs, and it can be particularly useful in the
presence of masks with non-trivial topology. Analytical modeling of
weak lensing statistics relies on an accurate modeling of the
statistics of underlying density distribution. We apply three
different formalisms to model the underlying dark-matter bispectrum:
the hierarchical ansatz, halo model and a fitting function based on
numerical simulations;  MFs resulting from each of these formalisms
are computed and compared. We investigate the extent to witch
late-time gravity-induced non-Gaussianity (to which weak lensing is
primarily sensitive) can be separated from primordial
non-Gaussianity and how this separation depends on source redshift
and angular scale.
\end{abstract}


\section{Introduction}

Since the first  measurements were published
\citep{BRE00,Wittman00,KWL00,Waerbeke00} there has been tremendous
progress in the field of weak gravitional lensing, regarding
analytical modelling, as well as technical specification and control
of systematics in observational surveys. Ongoing and planned weak
lensing surveys (see \citet{MuPhysRep08} for a review) such as the
CFHT{\footnote{http://www.cfht.hawai.edu/Sciences/CFHLS/}} Legacy
Survey, Pan-STARRS{\footnote{http://pan-starrs.ifa.hawai.edu/}}, the
Dark Energy Survey{\footnote{https://www.darkenergysurvey.org/}},
and, further in the future, the Large Synoptic Survey
Telescope{\footnote{http://www.lsst.org/llst\_home.shtml}},
JDEM{\footnote{http://jdem.gsfc.nasa.gov/}} and Euclid
{\footnote{http://http://sci.esa.int/euclid}} will map the
cosmological distribution of dark matter and probe the properties of
dark energy in unprecedented detail. Owing to the greater sky
coverage, tighter control on systematics and increased
number-density of source galaxies it will be soon possible to
extract higher-order statistics (i.e. beyond the two-point
correlation function), such as multispectra; see e.g. \cite{Pen03}.
Non-linearity induced by gravitational effects is generally used to
break the degeneracy between the amplitude of matter power spectrum
$\sigma_8$ and the matter density parameter $\Omega_M$; three-point
statistics such as the bispectrum (the three-point multispectrum)
are the best studied statistics for this purpose
\citep{Vil96,JainSeljak97}. Weak lensing can therefore play an
important role in breaking degeneracies, which makes it an ideal
complement to Cosmic Microwave Background (CMB) studies and studies
involving large scale structure (LSS) surveys.


Two-point statistics, principally the power spectrum, of density
perturbations remain the most frequently used statistical tool for
many cosmological studies. Weak lensing surveys probe the non-linear
regime and are therefore sensitive to non-Gaussian signatures which
can not be probed using two-point statistics. The statistics of
shear or convergence probe the statistics of underlying mass
distribution in an unbiased way
\citep{JSW00,MuJai01,Mu00,MuJa00,MuJai01,Valageas00,
MuVa05,VaMuBa04,VaMuBa05,TakadaWhite03,TakadaJain04}, but are very
sensitive to nonlinear evolution driven by gravitational clustering.
A number of analytical schemes, from perturbative calculations to
halo models have therefore been employed to model weak lensing
statistics \citet{Fry84,Schaeffer84, BerSch92,SzaSza93, SzaSza97,
MuBaMeSch99, MuCoMe99a, MuCoMe99b, MuMeCo99, MuCo00, MuCo02, MuCo03,
CooSeth02}). In addition to studying the statistics in projection on
the sky, they have also been studied in three dimensions using
photometric redshifts. It has been demonstrated that this approach
can tighten observational constraints on such quantities as the
neutrino mass and the dark energy equation of state
parameter\citep{Heav03,HRH00, HKT06, HKV07, Castro05, Kit08}.
Tomographic techniques have also been employed as an intermediate
strategy between projected surveys and 3D mapping
\citep{Hu99,TakadaJain04,TakadaJain03,Semboloni08}.

\begin{figure}
\begin{center}
{\epsfxsize=11. cm \epsfysize=5.4 cm {\epsfbox[27 432 585 713]{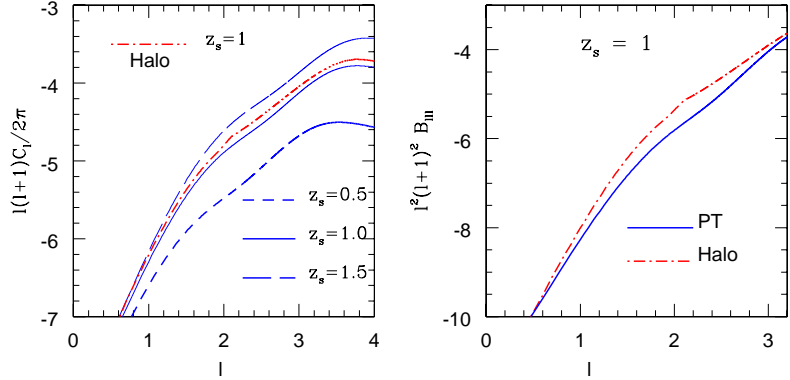}}}
\end{center}
\caption{The power spectrum is plotted as a function of the
harmonics $l$ in the left panel. A WMAP7 background cosmology is
used. The results are displayed for three different source redshifts
$z_s=1.5$, $1.0$ and $0.5$. The cosmological parameters are $\Gamma
= 0.1956$, $\Omega = 0.279$, $\Lambda = 0.721$ and $\sigma_8 =0.817$
respectively. The dot-dashed curve shows predictions from the halo
model for the same WMAP7 background cosmology and for the source
redshift $z_s=1$. The diagonal entries of the bi-spectrum are
plotted as a function of the harmonics $l$ in the right panel. The
results are for $z_s=1.0$. Two different approaches are persued in
computation of the bispectrum. The bispectrum results based on
extnsions of perturbation theory Eq.(\ref{eq:bispec_pert}) are
plotted using solid lines, the halo model predictions i.e.
Eq.(\ref{halocorr}) are shown using dashed lines.} \label{fig:S0}
\end{figure}


Minkowski Functionals (MFs) are morphological descriptors that are
commonly used in many cosmological contexts. They can be defined for
both  2D (projected) and  3D (redshift) data, and have  been used to
probe non-Gaussianity in CMB data \citep{Nat10,Komatsu03, Erik04},
weak lensing surveys \citep{MJ01,Sato01,Taru02} and galaxy surveys
\citep{GMD86,Co88,G89,Mel89,M92, Gott92,RGP92,Can98,Pk05,Hk08,HKM06,
HTS03, Hk02}. Unlike the multispectra, discussed above the
topological descriptors carry information of all orders (in a
statistical sense). In the context of CMB studies, the MFs are used
to probe primordial non-Gaussianity. For large scale structure
studies using projected or redshift galaxy surveys, the
non-Gaussianity probed is mainly that which is induced by gravity.
While galaxy surveys suffer from uncertainties relating to the
nature of galaxy  bias, weak lensing surveys will provide an
unbiased probe to probe the clustering of dark matter. The MFs will
be an important tool in this direction, along with other statistics
that can be used to probe non-Gaussianity to break the
parameter-space that are unavoidable when the  power spectrum alone
is used.


This paper is organized as follows. In \textsection2 we review the
formalism of Minkowski Functionals. In \textsection3 we link the
statistics of weak lensing convergence and the underlying density
distribution. In \textsection4 we introduce the concept of
generalized skew-spectra and show how these power spectra can be
used to study the Minkowski Functionals. In \textsection5 we review
the analytical models that are typically used for modelling of dark
matter clustering.

\section{Formalism}
\label{formalism}

The MFs are well-known morphological descriptors which are used in
the study of random fluctuation fields fields. Morphological
properties are defined to be those properties that remain invariant
under rotation and translation; see \cite{Hadwiger59} for a more
formal introduction. They are defined over an excursion set $\Sigma$
for a given threshold $\nu$. The three MFs that are defined for two
dimensional (2D) studies can be expressed, following the notations
of \citep{Hk08}, as:

\be
V_0(\nu) = \int_{\Sigma} da; \quad V_1(\nu) = {1 \over 4}\int_{\partial\Sigma} dl;
\quad V_2(\nu) = {1 \over 2\pi}\int_{\partial \Sigma} {\cal K} dl
\ee

\n Here $da$, $dl$ are surface are and line elements for the
excursion set $\Sigma$ and its boundary $\partial \Sigma$
respectively. The MFs $V_k(\nu)$ correspond to the area of the
excursion set $\Sigma$, the length of its boundary $\partial\Sigma$
as well as the integral of curvature $\cal K$ along its boundary
which is also related to the genus $g$ and hence the Euler
characteristics $\chi$.

In our analysis we will consider a smoothed random field
$\myf(\oh)$, with mean $\la \myf(\oh)\ra=0$ and variance $\sigma_0^2
= \la \myf^2(\oh) \ra$;  for the time being  $\myf$ is a generic 2D
weakly non-Gaussian random field defined on the sky although we will
introduce more specific examples later on. The spherical harmonic
decomposition, using $Y_{lm}(\oh)$ as basis functions, $\myf(\oh) =
\sum_{lm} \myf_{lm} Y_{lm}(\oh)$, can be used to define the power
spectrum $\myC_l$ using $\la \myf_{lm} \myf^*_{l'm'} \ra = \myC_l
\delta_{ll'}\delta_{mm'}$ which is a sufficient statistical
characterization of a Gaussian field. For a non-Gaussian field,
higher-order statistics such as the bi- or tri-spectrum can describe
the resulting mode-mode coupling. An alternative to this laborious
expansion in multispectra, topological measures such as the
Minkowski functionals can be employed to quantify deviations from
Gaussianity and it can be shown that the information content in both
descriptions is the same. At leading order the MFs can be
constructed completely from the knowledge of the bispectrum alone.
We will be studying the MFs defined over the surface of the
celestial sphere but equivalent results can be obtained in 3D using
Fourier decomposition (Munshi 2010, in preparation). The behaviour
of the MFs for a random Gaussian field is well known and is given by
Tomita's formula \citep{Tom86}. The MFs are denoted by $V_k(\nu)$
for a threshold $\nu= \myf/\sigma_0$, where $\sigma^2_0=\la \myf^2
\ra$ can be decomposed into two different contributions, Gaussian
($V^G_k(\nu)$) and non-Gaussian ($\delta V_k(\nu)$), i.e.  $V_k(\nu)
= V_k^G(\nu) + \delta V_k(\nu)$. From our perspective we will be
more interested in the non-Gaussian contribution, i.e. $\delta
V_k(\nu)$. We will further separate out an amplitude $A$ in the
expressions of both of these contributions which depend only on the
power spectrum of the perturbation through $\sigma_0$ and $\sigma_1$
(see e.g. \cite{Hk08}):

\ben
&& V^G_k(\nu) = A \exp \left ( -{\nu^2 \over 2}\right )  H_{k-1}; \quad\quad
\delta V_k(\nu) = A \exp \left ( -{\nu^2 \over 2}\right )
\left [ \delta V_k^{(2)}(\nu)\sigma_0 + \delta V_k^{(3)}(\nu)\sigma_0^2 + \delta V_k^{(4)}(\nu)\sigma_0^3 + \cdots \right ]  \\
&& \delta V_k^{(2)}(\nu) = \left [ \left \{  {1 \over 6} S^{(0)} H_{k+2}(\nu) + {k \over 3} S^{(1)} H_k(\nu) + {k(k-1) \over 6} S^{(2)} H_{k-2}(\nu)\right \} \right ];
\quad\quad
A = {1 \over (2\pi)^{(k+1)/2}} {\omega_2 \over \omega_{2-k}\omega_k}\left ( \sigma_1 \over \sqrt 2 \sigma_0 \right )^k.
\label{eq:v_k}
\een

\begin{figure}
\begin{center}
{\epsfxsize=6.5 cm \epsfysize=6.5 cm {\epsfbox[32 443 305 713]{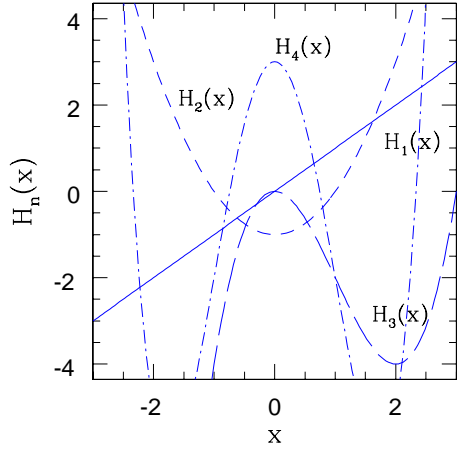}}}
\end{center}
\caption{The hermite polynomials that are used as a basis function for expanding the
Minkowski functional in the weakly non-gaussian limit. The plots show
$H_1(x)$,(solid) $H_2(x)$,(short-dashed) $H_3(x)$,(long-dashed) $H_4(x)$(dot-dashed) as a function of
the argument $x$ as depicted. }
\label{fig:S0}
\end{figure}

\n The constant $\omega_k$ introduced above is the volume of the
unit sphere in $k$ dimensions, i.e. $w_k = \pi^{k/2}/\Gamma(k/2+1)$;
in 2D we will only need $\omega_0=1$, $\omega_1=2$ and $\omega_2
=\pi$. The lowest-order Hermite polynoimals $H_k(\nu)$ are listed
below. As mentioned previously, the expressions consist of two
distinct contributions. The part which does not depend on the three
different skewness paramters $S^{(0)}, S^{(1)}, S^{(2)}$ signifies
the MFs for a Gaussian random field. The other contribution $\delta
V_k(\nu)$ represents the departure from the Gaussian statistics and
depends on the generalised skewness parameters defined in
Eq.(\ref{skewness_real_space}) and
Eq.(\ref{skewness_harmonic_space}). We have expanded the total
non-Gaussian contribution into perturbation series in $\sigma_0$.
While the lowest order terms $\delta V_k^{(2)}(\nu)$ are determined
by various one-point moments related to the bispectrum the next-to-
leading-order terms  $\delta V_k^{(3)}(\nu)$ are connected to
similar one-point moments related to trispectrum (also known as the
kurtosis). In projected surveys, even for relatively small angular
smoothing scales, the leading order terms are sufficient to describe
the non-Gaussian departures in the smoothed convergence field
$\kappa(\theta_s)$.

\ben
&& H_{-1}(\nu) = \sqrt{\pi \over 2} e^{\nu^2 / 2} {\rm erfc} \left (\nu \over \sqrt 2 \right ); \quad H_0(\nu) = 1, \quad H_1(\nu) = \nu, \nn \\
&& H_2(\nu)=\nu^2 -1, \quad H_3(\nu)=\nu^3 - 3\nu, \quad \quad H_4(\nu) = \nu^4 - 6\nu^2 + 3. \nn \\
&& H_n(\nu) = (-1)^n \exp \left ({ \nu^2 \over 2 } \right ) {d \over d\nu^n} \exp \left ({-\nu^2 \over 2 }\right )
\een

\n The various quantities $\sigma_j$ that appear in
Eq.(\ref{eq:v_k}) can be expressed in terms of the power spectra
$\myC_l$ and the shape of the observational beam $b_l$. The moment
$\sigma_0$ is a special case for  which $\sigma_0^2$ corresponds to
the variance. The quantities $\sigma_1$, $\sigma_2$ are natural
generalisations of variance, with increasing $j$ corresponding to
increase weight towards higher harmonics $\sigma_j^2 = {1 \over 4\pi
}\sum (2l+1) [l(l+1)]^j \myC_l b_l^2$. The variance that will mostly
be used are $\sigma_0^2 = \la \myf^2 \ra $ and $\sigma_1^2 = \la
(\nabla \myf)^2 \ra$.

The real-space expressions for the triplets of skewness $S^{(i)}$
are given below. These are natural generalisations of the ordinary
skewness $S^{0}$ that is used in many cosmological studies. They all
are cubic statistics but are constructed from different cubic
combinations:

\be S^{(0)} \equiv {S^{(\myf^3)} \over \sigma_0^4} = {\la \myf^3 \ra
\over \sigma_0^4}; \quad S^{(1)} \equiv-{3 \over
4}{S^{(\myf^2\nabla^2 \myf)} \over \sigma_0^2} = -{3 \over 4}{\la
\myf^2 \nabla^2 \myf \ra \over \sigma_0^2 \sigma_1^2}; \quad S^{(2)}
\equiv S^{(\nabla \myf \cdot \nabla \myf \nabla^2 \myf)} = -{3}{\la
(\nabla \myf).(\nabla \myf) (\nabla^2\myf) \ra  \over \sigma_1^4}.
\label{skewness_real_space} \ee

\n The expressions in the harmonic domain are more useful in the
context of CMB studies where we will be recovering them from a
masked sky using analytical tools that are commonly used for power
spectrum analysis. The skewness parameter $S^{(1)}$ is constructed
from the product field $\myf^2$ and $\nabla^2 f$, whereas skewness
parameter $S^{(2)}$ relies on the construction of $[\nabla
\myf\cdot\nabla \myf]$ and $\nabla^2 \myf$. By construction, the
skewness parameter $S^{(2)}$ has the highest weight at high $l$
modes and   $S^{(0)}$ has the lowest weights on high $l$ modes. The
expressions in terms of the bispectrum $B_{l_1l_2l_3}$ (see
Eq.(\ref{eq:bispec}) for defintion) take the following form (see
e.g. \cite{Hk08}):

\ben
&& S^{(\myf^3)} = {1 \over 4 \pi} \sum_{l_i} \myB_{l_1l_2l_3}I_{l_1l_2l_3}  \\
&& S^{(\myf^2\nabla^2\myf)} = -{1 \over 12 \pi} \sum_{l_i}{} \Big [{l_1(l_1+1)+ l_2(l_2+1)+ l_3(l_3+1)} \Big ] \myB_{l_1l_2l_3}I_{l_1l_2l_3}  \\
&& S^{(\nabla \myf\cdot\nabla \myf \nabla^2 \myf)} = {1 \over 4 \pi} \sum_{l_i}
{}\Big [ [l_1(l_1+1)+l_2(l_2+1) - l_3(l_3+1)]l_3(l_3+1) + {\rm cyc.perm.} \Big ]
 \myB_{l_1l_2l_3}I_{l_1l_2l_3}W_{l_1}W_{l_2}W_{l_3}\\
&& I_{l_1l_2l_3} = \sqrt{(2l_1+1)(2l_2+1)(2l_3+1) \over 4 \pi }\left ( \begin{array}{ c c c }
     l_1 & l_2 & l_3 \\
     0 & 0 & 0
  \end{array} \right).
\label{skewness_harmonic_space}
\een

\n The bispectrum $B_{l_1l_2l_3}$ used here defines the three-point
correlation function in the harmonic domain. A reduced bispectrum
$b_{l_1l_2l_3}$ can also be defined which can directly be linked to
the {\em flat-sky} expressions.

\be
\la \kappa_{l_1m_1}\kappa_{l_2m_2}\kappa_{l_3m_3} \ra_c = \left ( \begin{array}{ c c c }
     l_1 & l_2 & l_3 \\
     m_1 & m_2 & m_3
  \end{array} \right) \myB_{l_1l_2l_3}; \quad\quad \myB_{l_1l_2l_3} =
  I_{l_1l_2l_3}b_{l_1l_2l_3}.
\label{eq:bispec} \ee The expressions for the MFs in
Eq.(\ref{eq:v_k}) depend on the one-point cumulants $S^{(i)}$.
However it is possible to define power spectra associated with each
of these skewnesses following a procedure developed in
\cite{MuHe10}. This will mean we can also associate a power spectrum
with $V_k^{(3)}$ which will generalize the concept of MFs in a
scale-dependent way. The power spectrum that we associate with MFs
will have the same correspondence with various skew-spectra
$S^{(i)}_l$ as the MFs have with one-point cumulants or $S^{(0)}$.
The power spectra so defined will however have more power to
distinguish various models of non-Gaussianity. This is one of the
main motivations behind generalising the concept of MFs, each of
which is a number, to a power spectrum, which contains scale
information.

The series expansion for the MFs can be extended beyond the level of
the bispectrum; the next-to-leading-order corrections terms are
related to trispectra of the original fields and various derivatives
constructed from them using differential operations such as
$\nabla\cdot\nabla, \nabla^2$. These corrections are expected to be
sub-dominant in the context of CMB studies for the entire range of
angular scales being probed.

The results here correspond to analysis of convergence maps, which
are spin-$0$ objects. It is possible to extend these results to
spin-2 fields. Such results will be interesting for analysis of weak
lensing shear and flexions, but a detailed analysis will be
presented elsewhere.

\section{Convergence power spectrum and bispectrum}

The convergence $\kappa(\oh,r_s)$ can be treated as a line-of-sight
projection of the density contrast $\delta(\oh,r)$ along the
direction $\oh$ ($r$ is comoving radial distance) out to a source
redshift $z_s(r_s)$ with a redshift-dependent weight function
$\omega(r,r_s)$:

\be
\kappa(\oh,r_s) = \int_0^{r_s} \,dr\, w(r,r_s)\,\delta(\oh,r); \quad
\omega(r,r_s) = {3 \over 2a}{H_0^2 \over c^2}\Omega_M {d_A(r-r_s)\over d_A(r)d_A(r_s)};
\quad r_s = {\rm min}(r_1,r_2).
\ee

\n The weight functions $\omega(r)$ for weak lensing depend on the
angular diameter distance $d_A(r)$, Hubble constant $H_0$, matter
density parameter $\Omega_M$ and the scale factor of the Universe
$a=1/(1+z)$ at a redshift $z$. The angular diameter distance
$d_A(r)$ is linked to the total matter content $\Omega_0$ and the
Hubble constant $H_0$, i.e. $d_A(r) = {\rm K}^{-1/2}\sin({\rm
K}^{-1/2}r),\; {\rm K}^{-1/2}\sinh((-{\rm K})^{-1/2}r),\; r$ for
open, closed and flat Universes; here ${\rm K} = (\Omega_0-1)H_0^2$.
We will consider the projected cross-power spectra $C_l$ that depend
on two different redshift $z_{1}$ and $z_{2}$ which is a function of
the underlying matter power spectra $P_{\delta}(k,r)$ which, in the
small-angle approximation \citep{Limb54}, can be expressed as
\citep{Kaiser92}:

\be
\myC_l =  \int_0^{r_s}\,dr\,
\,{w^2(r;r_{s})\over d^2_A(r)} \,P_{\delta}\left ({l \over d_A(r)}, r \right ); \quad
\label{ps}
\ee

Analytical modelling of the convergence bispectrum ${\cal B}$
depends on modelling of the underlying matter bispectrum
$B_{\delta}$:

\be
{\cal B}_{l_1l_2l_3} =
I_{l_1l_2l_3}\int_0^{r_s} dr \,
{w^3(r,r_s) \over d^4_A(r)} \, B_{\delta}\left ({l_1 \over d_A(r)},{l_2 \over d_A(r)},{l_3 \over d_A(r)}
\right ); \quad
\label{bips}
\ee

\n we will discuss the analytical models we use to construct
$B_{\delta}$ in later sections.This equation can also be used to express the reduced bispectrum $b_{l_1l_2l_3}$ introduced before in Eq.(\ref{eq:bispec}). Estimation of individual modes of
the bispectrum defined by specific choice of the triplets
$(l_1,l_2,l_3)$ is difficult when the data are noisy, but it is
possible to extract the cross-correlation of product maps
$\kappa^2(\oh)$ against $\kappa(\oh)$. If we denote the harmonics of
the product map
 $\kappa^2(\oh)$ as $\kappa_{lm}^{(2)} = \int\, d\oh\, Y^*_{lm}(\oh)\kappa^2(\oh)$
and similarly $\kappa_{lm}^{} = \int \,d\oh\, Y^*_{lm}(\oh)
\kappa(\oh);$ then the associated power spectrum  is constructed as
$\myC_l^{(2,1)} = {1 \over 2l+1}\sum_l {\rm Re}
(\kappa_{lm}^{(2)}\kappa_{lm}^{(1)})$ is called the skew-spectrum
\citep{Cooray01}. We will next generalize the concept of
skew-spectrum and introduce a set of generalized skew-spectrum that
can be used to construct the Minkowski Functionals at the lowest
level of non-Gaussianity.

\section{The triplets of skew-spectra and lowest-order corrections to Gaussian MFs}

\n The skew-spectra are cubic statistics that are constructed by
cross-correlating two different fields. One of the field used is a
composite field typically a  product of two maps either in its
original form or constructed by means of relevant differential
operations. The second field will typically be a single field but
may be constructed by applying various differential operators. All
three skewnesses contribute to the three MFs that we will consider
in 2D.

The first of the skew-spectra was studied by \citep{Cooray01} and
was later generalized by \cite{MuHe10} and is related to commonly
used skewness. The skewness in this case is constructed by
cross-correlating the squared map $[\myf^2(\oh)]$ with the original
map $[\myf(\oh)]$.  The second skew-spectrum is constructed by
cross-correlating the squared map $[\myf^2(\oh)]$ against
$[\nabla^2\myf(\oh)]$. Analogously, the third skew-spectrum
represents the cross-spectra that can be constructed using $[\nabla
\myf(\oh)\cdot\nabla \myf(\oh)]$ and $[\nabla^2 \myf(\oh)]$ maps.

\ben
&& S_l^{(0)} \equiv {1 \over 12 \pi \sigma_0^4}S_l^{(\myf^2,\myf)} \equiv {1 \over 12 \pi \sigma_0^4}{1 \over 2l+1}\sum_m
{\rm Real}([\myf]_{lm}[\myf^2]^*_{lm})  ={1 \over 12 \pi \sigma_0^4} \sum_{l_1l_2} \myB_{ll_1l_2}J_{ll_1l_2}W_{l}W_{l_1}W_{l_2}  \\
&& S_l^{(1)} \equiv {1 \over 16 \pi \sigma_0^2\sigma_1^2} S_l^{(\myf^2,\nabla^2 \myf)}
\equiv {1 \over 16 \pi \sigma_0^2\sigma_1^2}{1 \over 2l+1}\sum_m {\rm Real}([\nabla^2 \myf]_{lm}[\myf^2]^*_{lm}) \nn \\
&&\quad\quad = {1 \over 16 \pi \sigma_0^2\sigma_1^2} \sum_{l_i} \Big [{l(l+1)+ l_1(l_1+1)+ l_2(l_2+1)} \Big ] \myB_{ll_1l_2}J_{ll_1l_2}
W_{l}W_{l_1}W_{l_2}  \\
&& S_l^{(2)} \equiv {1 \over 8 \pi \sigma_1^4}S_l^{(\nabla \myf\cdot\nabla \myf, \nabla^2\myf)} \equiv {1 \over 8 \pi \sigma_1^4} {1 \over 2l+1}\sum_m
{\rm Real}([\nabla \myf \cdot \nabla \myf]_{lm}[\nabla^2 \myf]^*_{lm}) \nn \\
&&\quad\quad ={1 \over 8 \pi \sigma_1^4} \sum_{l_i}
{}\Big [ [l(l+1)+l_1(l_1+1) - l_2(l_2+1)]l_2(l_2+1) + {\rm cyc.perm.} \Big ]
\myB_{ll_1l_2}J_{ll_1l_2}W_{l}W_{l_1}W_{l_2}\\
&& J_{l_1l_2l_3} \equiv {I_{l_1l_2l_3} \over 2l_3+1} = \sqrt{(2l_1+1)(2l_2+1) \over (2l_3+1) 4 \pi }\left ( \begin{array}{ c c c }
     l_1 & l_2 & l_3 \\
     0 & 0 & 0
  \end{array} \right). \\
&& S^{(i)} = \sum_{l}(2l+1)S^{(i)}_l\\
&& \sigma_j^2 = {1 \over 4\pi}\sum_l (2l+1)[l(l+1)]^j \myC_l W_l^2
\label{skew_spectra}
\een

\n This set of equations constitutes one of the main results of this
paper. The matrices here denote the Wigner-$3j$ symbols, $W_l$
represents the smoothing window, e.g. a top hat, Gaussian or some
form of compensated filter. Each of these spectra probes the same
bispectrum $B_{ll_1l_2}$ with different weights for individual
triplets of modes that specify the bispectrum $(l,l_1l_2)$ and
define a triangle in the harmonic domain. The skew spectra is summed
over all possible configurations of the bispectrum keeping one of
its sides at a fixed $l$. For each individual choice of $l$ we can
compute the skew-spectrum  $S_l^{(i)}$ relatively straightforwardly.
by constructing the relevant maps in real space (either by algebraic
or differential operation) and then cross-correlating them in the
multipole domain. Issues related to mask and noise will be dealt
with in later sections, where we will show that, even in the
presence of a mask, the computed skew spectra can be inverted to
give a unbiased estimate of all-sky skew-spectra. Presence of noise
will only affect the scatter. We have explicitly displayed the
experimental beam $b_l$ in all our expressions.

To derive the above expressions, we first express the spherical
harmonic expansion of the fields $[\nabla^2 \myf(\oh)]$, $[\nabla
\myf(\oh) \cdot \nabla \myf(\oh)]$ and $[\myf^2(\oh)]$ in terms of
the harmonics of the original fields $\myf_{lm}$. These expressions
involve the 3j functions as well as factors that depend on various
$l_i$ dependent weight factors.

\ben
&& [\nabla^2 \myf(\oh)]_{lm} = \int ~d\oh~ Y^*_{lm}(\oh)~ [\nabla^2 \myf(\oh)] = -l(l+1)\myf_{lm} \nn \\
&&[\myf^2(\oh)]_{lm} = \int~d\oh~Y^*_{lm}(\oh)~[\myf^2(\oh)] = \sum_{l_im_i} (-1)^m\myf_{l_1m_1} \myf_{l_2m_2}I_{l_1l_2l}
\left ( \begin{array}{ c c c }
     l_1 & l_2 & l\\
     m_1 & m_2 & -m
  \end{array} \right). \nn \\
&& [\nabla \myf(\oh) \cdot \nabla \myf(\oh)]_{lm}  = \int d\oh Y^*_{lm}(\oh)[\nabla \myf(\oh) \cdot \nabla \myf(\oh)] =
\sum_{l_im_i}  \myf_{l_1m_1} \myf_{l_2m_2} \int~d\oh~Y^*_{lm}(\oh)~ [\nabla Y_{l_1m_2}(\oh) \cdot \nabla Y_{l_2m_2}(\oh)] \\
&& \quad\quad\quad  = {1 \over 3}  \sum_{l_im_i}[l_1(l_1+1)+l_2(l_2+1)-l(l+1)] \int d\oh Y^*_{lm}(\oh) Y_{l_1m_1}(\oh)Y_{l_2m_2}(\oh) \nn \\
&& \quad\quad\quad  = {1 \over 3} \sum_{l_im_i}(-1)^m[l_1(l_1+1)+l_2(l_2+1)-l(l+1)]
\myf_{l_1m_1} \myf_{l_2m_2} I_{l_1l_2l}
\left ( \begin{array}{ c c c }
     l_1 & l_2 & l \\
     m_1 & m_2 & -m
  \end{array} \right).
\label{eq.harmonics}
\een

\n
We can define the power spectrum associated with the MFs through the following third order expression:

\be
V_k^{(3)} = \sum_l[V_k]_l(2l+1) = {1 \over 6 } \sum_l (2l+1) \left \{ S^{(0)}_l H_k(\nu) + {k \over 3 } S^{(1)}_l H_{k-1}(\nu) +
{k(k-1) \over 6 } S^{(2)}_l H_{k-2}(\nu) + \cdots \right \}.
\ee

\n The three skewnesses thus define triplets of Minkowski
Functionals. At the level of two-point statistics, in the harmonic
domain, we have three power-spectra associated with MF $V_k^{(3)}$
that depend on the three skew-spectra we have defined. We will show
later in this paper that the fourth order correction terms too have
a similar form with an additional monopole contribution that can be
computed from the lower order one-point terms in a similar way as
the three skewness defined here. The result presented here is
important and implies that we can study the contributions to each of
the MFs $v_k(\nu)$ as a function of harmonic mode $l$. This is
especially significant result as various form of non-Gaussianity
will have different $l$ dependence and so can potentially be
distinguished from each other using this approach. The ordinary MFs
add contributions from all individual $l$ modes and hence have less
power in differentiating various contributing sources of
non-Gaussianity. This is one of main motivations to extend the
concept of MFs (single numbers) to one-dimensional objects similar
to power spectrum.

It is worth pointing out that the skewness and generalized skewness
parameters are relatively insensitive to the background cosmology
but quite sensitive to the underlying model of non-Gaussianity. The
main dependence on cosmology typically results from the
normalization coefficients such as $\sigma_0$ and $\sigma_1$ which
are determined the power spectrum of the convergence $\kappa$.

In real space the skew-spectra can be defined through these
correlation functions:

\be
S^{(0)}(\oh_1,\oh_2) \equiv  \la \kappa^2(\oh_1))\kappa(\oh_2) \ra;
\quad S^{(1)}(\oh_1,\oh_2) \equiv \la \kappa^2(\oh_1) \nabla^2\kappa(\oh_2)\ra;
\quad S^{(2)}(\oh_1,\oh_2) \equiv \la\nabla \kappa(\oh_1)\cdot \nabla \kappa(\oh_1) \nabla^2\kappa(\oh_2) \ra
\label{eq:cumu_corr}
\ee

\n Although we have adopted a harmonic approach these correlations
can equivalently be used to probe MFs especially for smaller
surveys.

\section{Modelling the primordial and gravity-induced bispectrum}

\n  s  It is clear that we need accurate analytical modeling of dark
matter clustering for prediction of weak lensing statistics, but in
general there is no definitive analytical theory for handling
gravitational clustering in the highly nonlinear regime. On larger
scales, where the density field is only weakly nonlinear,
perturbative treatments are known to be valid. For a
phenomenological statistical description of dark matter clustering
in collapsed objects on nonlinear scales, typically the halo model
\citep{CooSeth02} is used.  We will be using the Halo Model in our
study, but  an alternative to the Halo Model approach on small
scales is to employ various \emph{ansatze} which trace their origin
to field-theoretic techniques. Here we provide a quick summary of
some of the analytical prescriptions that can be used to model
non-linear clustering. We will also provide a brief description of
various models of primordial non-Gaussianity arising from variants
of the inflationary universe scenario.

\subsection{Hierarchical ansatz}

The hierarchical \emph{ansatz} has been used for many weak lensing
related work, where the higher-order correlation functions are
constructed from the two-point correlation functions. Assuming a
tree model for the matter correlation hierarchy (typically used in
the highly non-linear regime) one can write the most general case,
the $N$ point correlation function, $\langle \delta({\bf r_1}) \dots
\delta({\bf r_n})\rangle_c = \xi_N^\delta({\bf r}_1, \dots, {\bf
r}_n)$ as a product of two-point correlation functions $\la
\delta(\br_i)\delta(\br_j)\ra_c = \xi_2^\delta(|\br_i-\br_j|)$
\citep{Bernardreview02}. Equivalently, in the Fourier domain, the
multispectra can be written as products of the matter power spectrum
$P\ad(k_1)$. The temporal dependence is implicit here.

\begin{equation}
\xi_N({\bf r_1}, \dots, {\bf r_n}) \equiv \langle \delta({\bf r_1})
\dots \delta({\bf r_n})\rangle_c = \sum_{\alpha,{\rm N=trees}}
Q_{N,\alpha}\sum_{\rm labellings} \displaystyle \prod_{\rm edges
(i,j)}^{(N-1)} \xi_2(|\br_i-\br_j|).
\end{equation}

\n It is very interesting to note that a similar hierarchy develops
in the quasi-linear regime at tree-level in the limiting case of
vanishing variance, except that the hierarchical amplitudes become
shape-dependent in such a case. These kernels are also used to
relate the halo-halo correlation hierarchy to the underlying mass
correlation hierarchy. Nevertheless there are indications from
numerical simulations that these amplitudes become
configuration-independent again as has been shown by high resolution
studies for the lowest order case $Q_3 = Q$ \citep{Scocci98,
Bernardreview02}. See \cite{Waerbeke01} for related discussion about
use of perturbation theory results in intermediate scales. In
Fourier space, however, such an ansatz means that the entire
hierarchy of multi-spectra can be written in terms of sums of
products of power spectra with different amplitudes $Q_{N,\alpha}$
etc. The power spectrum is defined through $\langle
\delta(\bk_1)\delta(\bk_2) \rangle_c  = (2\pi)^3 \delta_{3D}({\bf
k}_{12})P^{\delta}_{nl}(k_1)$ . Similarly, the bispectrum and
trispectrum are defined through the following expressions $\langle
\delta(\bk_1) \delta(\bk_2)\delta(\bk_3) \rangle_c = (2\pi)^3
\delta_{3D}(\bk_{123})B^\delta(\bk_1,\bk_2,\bk_3)$ and $\langle
\delta(\bk_1) \cdots \delta(\bk_4) \rangle_c = (2\pi)^3
\delta_{3D}(\bk_{1234})T^\delta(\bk_1,\bk_2,\bk_3,\bk_4)$. The
subscript $c$ here represents the connected part of the spectrum and
$\bk_{i_1\dots i_n} = \bk_{i_1} + \dots + \bk_{i_n}$. The Dirac
delta functions $\delta_{3D}$ ensure translation invariance at each
vertex representing the multi-spectrum.

\begin{eqnarray}
&& B^\delta(\bk_1,\bk_2, \bk_3)_{\sum \bk_i=0} =
Q_3[P^\delta_{nl}(k_1)P^\delta_{nl}(k_2)+ P^\delta_{nl}(k_1)P^\delta_{nl}(k_3)+ P^\delta_{nl}(k_2)P^\delta(k_3)] \\
\label{eq:hier_bispec}
&& T^\delta(\bk_1,\bk_2,\bk_3,\bk_4)_{\sum \bk_i=0} = R_a[P^\delta_{nl}(k_1)P^\delta_{nl}(k_2)P^\delta_{lin}(k_3)
+ cyc.perm.]+ R_b[P^\delta_{nl}(k_1)P^\delta_{nl}(|\bk_{12}|)P^\delta_{nl}(|\bk_{123}|)+ cyc.perm.].
\end{eqnarray}

\noindent Different hierarchical models differ in the way numerical
values are allocated to the different amplitudes. \citet{BerSch92}
considered ``snake'', ``hybrid'' and ``star'' diagrams with
differing amplitudes at various order. A new ``star''  appears at
each order. higher-order ''snakes'' or ``hybrid'' diagrams are built
from lower-order ``star'' diagrams. In models where we only have
only star diagrams \citep{VaMuBa04}, the expressions for the
trispectrum takes the following form:
$T^\delta(\bk_1,\bk_2,\bk_3,\bk_4)_{\sum \bk_i=0}  =
Q_{4}[P^\delta(k_1)P^\delta(k_2)P^\delta(k_3) + cyc.perm.]$.
Following \cite{VaMuBa04} we will call these models ``stellar
models''. Indeed it is also possible to use perturbative
calculations which are however valid only at large scales. While we
still do not have an exact description of the non-linear clustering
of a self-gravitating medium in a cosmological scenario, these
approaches do capture some of the salient features of gravitational
clustering in the highly non-linear regime and have been tested
extensively against numerical simulation in 2D statistics of
convergence of shear \citep{VaMuBa04}. These models have also been
used for modelling the covariance of lower-order cumulants
\citep{MuVa05}.

\begin{figure}
\begin{center}
{\epsfxsize=11 cm \epsfysize=5.5 cm {\epsfbox[32 416 587 709]{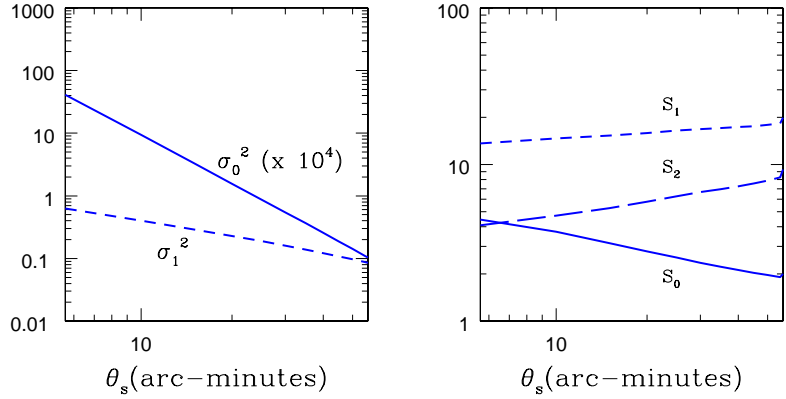}}}
\end{center}
\caption{The moments $\sigma^2_0(\theta_s)$ and
$\sigma^2_1(\theta_s)$ (left panel) and the skewness paramters
$S^{(0)}(\theta_s)$, $S^{(1)}(\theta_s)$ and $S^{(2)}(\theta_s)$
(right panel) are plotted for source redshift $z_s=1$ as a function
of smoothing angular scale $\theta_s$  see Eq.(\ref{skew_spectra})
for definitions of $\sigma_i$ and $S^{(n)}$. The underlying
cosmology is assumed to be that of WMAP7. A top hat window is
assumed for both of this plot. The resolution is fixed at $l_{max} =
4000$. The underlying modelling of the convergence bispectrum ${\cal
B}_{l_1l_2l_3}$ depends on modelling of matter bispectrum
$B_{l_1l_2l_3}$. The specific model for the underlying model that
was used for this plot is based on perturbative results and its
extrapolation to highly non-linear regime; see text for more
details. The skewness paramaters are defined in
Eq.(\ref{skewness_harmonic_space}). The parameters $\sigma_j$ are
defined in Eq.(\ref{skew_spectra}). } \label{fig:moments}
\end{figure}

\begin{figure}
\begin{center}
{\epsfxsize=11. cm \epsfysize=5.5 cm {\epsfbox[32 418 582 713]{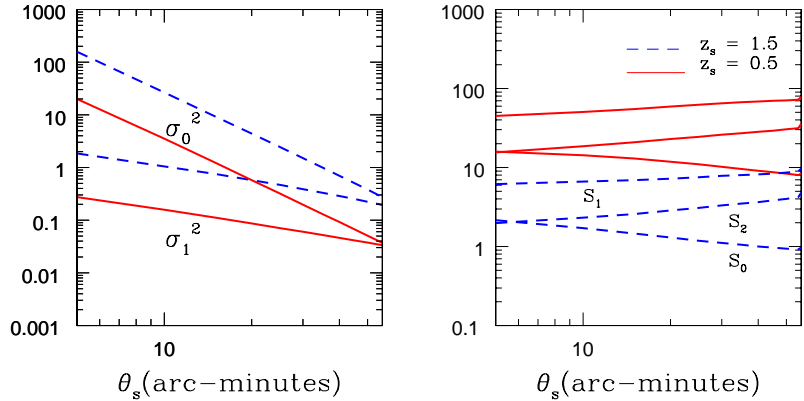}}}
\end{center}
\caption{Same as previous figure but for redshift $z_s=0.5$ and $z_s=1.5$ as indicated. The skewness paramaters are defined in
Eq.(\ref{skewness_harmonic_space}). The parameters $\sigma^2_j$ are defined in Eq.(\ref{skew_spectra}). For a given
angular smoothing the skewness parameters increase with redshift. The cosmological parameters of the background
cosmology is that of WMAP7. The variance parameters $\sigma^2_j$ increase with redshift however the skewness parameters
show an increasing trend. }
\label{fig:snz}
\end{figure}

The statistics of the projected convergence field can be constructed
using a suitable defined variable $\eta = (\kappa - \kappa_{\rm
min})/ \kappa_{min}$ where $\kappa(\oh,r_s) = -\int_0^{r_s} \,dr\,
w(r,r_s)$. The  variable $\eta$ follows the same statistics as the
density parameter $\delta$ and  under some simplifying assumptions
and using hierarchical ansatz it can be shown that $S^{(0)}=
S_3^{\delta}/ \eta$ and similar results also hold at higher order
i.e. $K^{(0)}= K_4^{\delta}/\eta^2$. The overall dependance on the
cosmology is absorbed in the definition of $\eta$ and the skewness
$S_3^{\delta} = 3Q$, kurtosis $K_4^{\delta} = 4R_a+ 12R_b \sim
16Q_4$ parameters, defined in terms of the hierarchical amplitudes,
$Q_3$ and $R_a, R_b$ are insensitive to the background cosmology
\citep{MuJai01,Mu00,MuCo00}.

\subsection{Halo Model}

The Halo Model relies on  a phenomenological model for the
clustering of halos and predictions from perturbative calculations
on large scales to model the non-linear correlation functions. The
halo over-density at a given position ${\bf x}$, $\delta^h({\bf x},
M; z)$ can be related to the underlying density contrast
$\delta({\bf x}, z)$ by a Taylor expansion \citep{Mo97}.

\be
\delta^h({\bf x}, M; z) = b_1(M;z) \delta({\bf x},z) + {1 \over 2} b_2(M,z) \delta^2({\bf x},z) + \dots
\ee

\n The expansion coefficients are functions of the threshold $\nu_c
= {\delta_c / \sigma(M,z)}$. Here $\delta_c$ is the threshold for a
spherical over-density to collapse and $\sigma(M,z)$ is the {\em
rms} fluctuation within a top hat filter. The halo model
incorporates  perturbative aspects of gravitational dynamics by
using it to model the halo-halo correlation hierarchy; the nonlinear
features of this take direct contributions from the halo profile.
The total power spectrum $P^t(k)$ at non-linear scale can be written
as

\be P^{1h} = I_2^0(k,k); \qquad P^{2h}(k) = [I_1^1(k)]^2 P(k);
\qquad P^t = P^{2h}(k) + P^{1h}(k) \label{eq:halo_ps} \ee

\citep{S00}. The minimum halo mas that we consider in our
calculation is $10^3 M_{\sun}$ and the maximum is $10^{16}M_{\sun}$.
more massive halos do not contribute significantly owing to their
low abundance. The bispectrum involves terms from one, two or three
halo contributions and the total can be written as

\beqa
&& B^t(k_1,k_2,k_3) = B^{3h}(k_1,k_2,k_3)+ B^{2h}(k_1,k_2,k_3)+ B^{1h}(k_1,k_2,k_3); \label{halocorr1} \\
&& B^{1h} = I_3^0(k_1,k_2,k_3); \qquad B^{2h}(k_1,k_2,k_3) = I_2^1(k_1,k_2)I_1^0(k_3)P(k_3) +cyc.perm.;
\label{halocorr2}  \\
&& B^{3h}(k_1,k_2,k_3) = [2J(k_1,k_2,k_3)I_1^1(k_3)
+ I_1^2(k_3)] I_1^1(k_1)I_1^1(k_2)P(k_1)P(k_2)+ cyc.perm.
\label{halocorr}
\eeqa

\begin{figure}
\begin{center}
{\epsfxsize=17. cm \epsfysize=5.5 cm {\epsfbox[32 533 582 713]{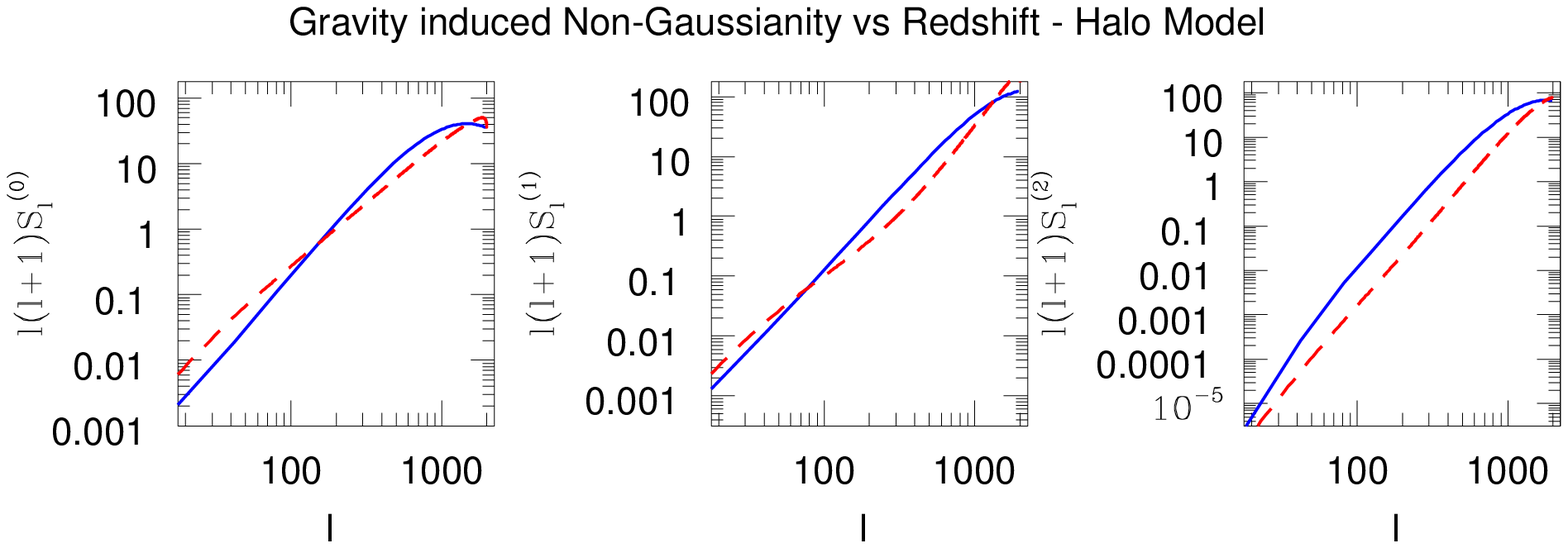}}}
\end{center}
\caption{The Halo Model is used to predict the skew spectra
$S_l^{(0)}$ (left panel), $S_l^{(1)}$ (middle panel) and $S_l^{(2)}$
(right panel). The source redshift is unity. The underlying
background cosmology is that of WMAP7. No smoothing window was
assumed. A sharp cutoff at $l_{max} = 2000$ was used for this
calculations. Halos in the mass range of $10^{3}M_{\sun} -
10^{16}M_{\sun}$ were include in this calculation. The halo model
expression for the bispectrum is defined in Eq.(\ref{halocorr1}) -
Eq.(\ref{halocorr}). The dashed lines correspond to the analytical
model prescribed in Eq.(\ref{eq:bispec_pert}).} \label{fig:halo}
\end{figure}

\n The kernel $J(k_1,k_2,k_3)$ is derived using second-order
perturbation theory \citep{Fry84,Bouchet92} and he integrals
$I_{\mu}^{\beta}$ can be expressed in terms of the Fourier transform
of halo profile (assumed to be an NFW \citep{NFW96}:

\be I^{\beta}_{\mu}(k_1,k_2,\dots, k_\mu;z) = \int dM \left ({M
\over \rho_b} \right )^\mu {dn(m,z) \over dM } b_{\beta}(M)
y(k_1,M)\dots y(k_\mu,M); \qquad y(k,M) = {1 \over
M}\int_{{0}}^{r_v} dr 4\pi r^2 \rho(r,M) \left [ {\sin(kr) \over kr
}\right ] \label{eq:gen_I} \ee The mass function is assumed to be
given by the Press-Schechter form \citep{PS74}. The results are
obtained by using Eq.(\ref{halocorr1})-Eq.(\ref{halocorr}) in
 Eq.(\ref{ps}) and Eq.(\ref{bips}). The convergence power
spectra and bispectra thus computed are then inserted in
Eq.(\ref{skew_spectra}).

\begin{figure}
\begin{center}
{\epsfxsize=16. cm \epsfysize=5.6 cm {\epsfbox[36 520 590 735]{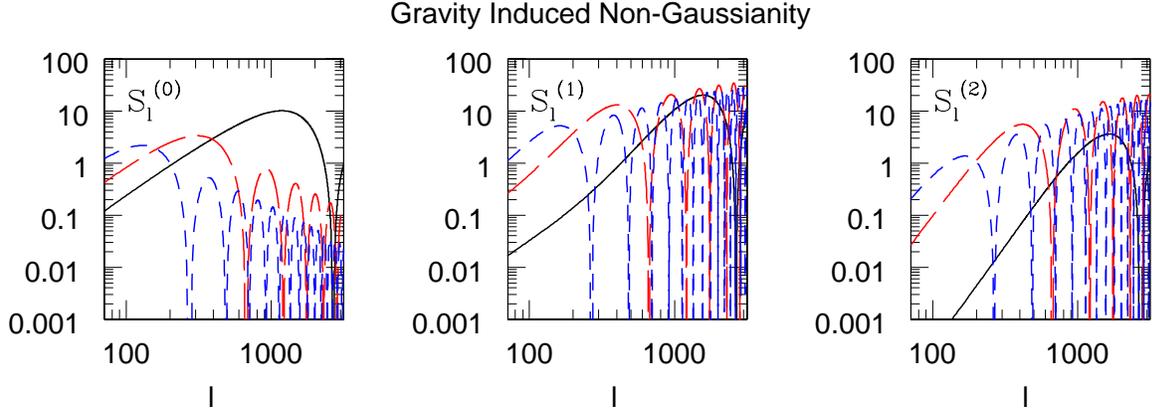}}}
\end{center}
\caption{The skew-spectra $S^{(0)}_l, S^{(1)}_l$ and $S^{(2)}_l$ are plotted for
$z_s=1$ as a function of wave number $l$. The underlying cosmology is that of WMAP7.
A tophat window is assumed. Various curves correspond to different smoothing angular scales
as indicated. The resolution is fixed at $l_{max}=4000$. The smoothing angular scales
considered are $\theta_s=5'$ (solid lines), $\theta_s=25'$ (long-dashed lines) and $\theta_s=55'$ (short-dashed lines) respectively. The skew-spectra
are defined in Eq.(\ref{skew_spectra}). The underlying bispectrum is constructed using the analytica model
prescribed in Eq.(\ref{eq:bispec_pert}). It is interesting to note that at smaller $l$ the skew-spectra
with larger smoothing angular scales $\theta_s$ dominates. However smaller smoothing angular scales dominates
at higher $l$ resulting in a higher values of the corresponding one-point skewness parameters. }
\label{fig:Sl}
\end{figure}

\begin{figure}
\begin{center}
{\epsfxsize=16. cm \epsfysize=5.6 cm {\epsfbox[36 520 590 735]{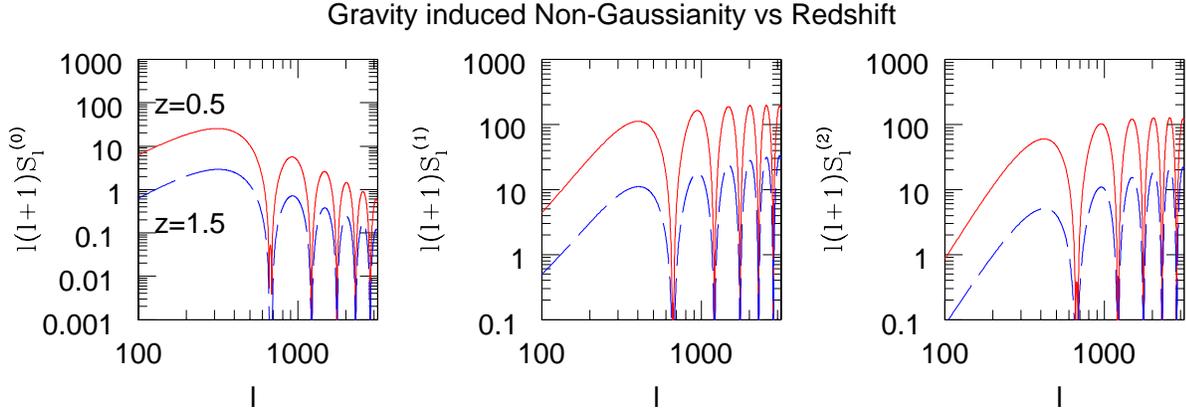}}}
\end{center}
\caption{The skew-spectra $S^{(0)}_l, S^{(1)}_l$ and $S^{(2)}_l$ are plotted for
$z_s=0.5$ (solid line) and $z_s=1.5$ (dashed line) as a function of wave number $l$. The underlying cosmology is that of WMAP7.
A tophat window is assumed. Various curves correspond to different smoothing angular scales
as indicated. The resolution is fixed at $l_{max}=4000$. The smoothing angular scale is fixed at
$\theta_s=25'$. Notice that use of broader window not only removed power at smaller angular scale,
it also changes the overall normalisation of the skew-spectra. The skew-spectra for any specific smoothing angular
scales increases with lowering of the source redshift. This is due to the fact the PDF of convergence for
higher redshift is more Gaussian than at a lower redshift. At a lower redshift the highly evolved large scale
structure results in higher departure of the convergence statistics from Gaussianity.}
\label{fig:Sl_red}
\end{figure}

Results from the Halo Model analysis are plotted for the
skew-spectra $S_l^{(0)}$ (left panel), $S_l^{(1)}$ (middle panel)
and $S_l^{(2)}$ (right panel). The source redshift is fixed at
unity. The underlying background cosmology is that of WMAP7. No
smoothing window was assumed. A sharp cutoff at $l_{max} = 2000$ was
used for this calculations. As mentioned, halos in the mass range of
$10^{3}M_{\sun} - 10^{16}M_{\sun}$ were included in this
calculation. The halo model expression for the bispectrum is defined
in Eq.(\ref{halocorr1}) - Eq.(\ref{halocorr}).

\subsection{Perturbative calculations in the quasi-linear regime and their extensions}
\label{pert}

\begin{figure}
\begin{center}
{\epsfxsize=16. cm \epsfysize=5.6 cm {\epsfbox[21 528 590 715]{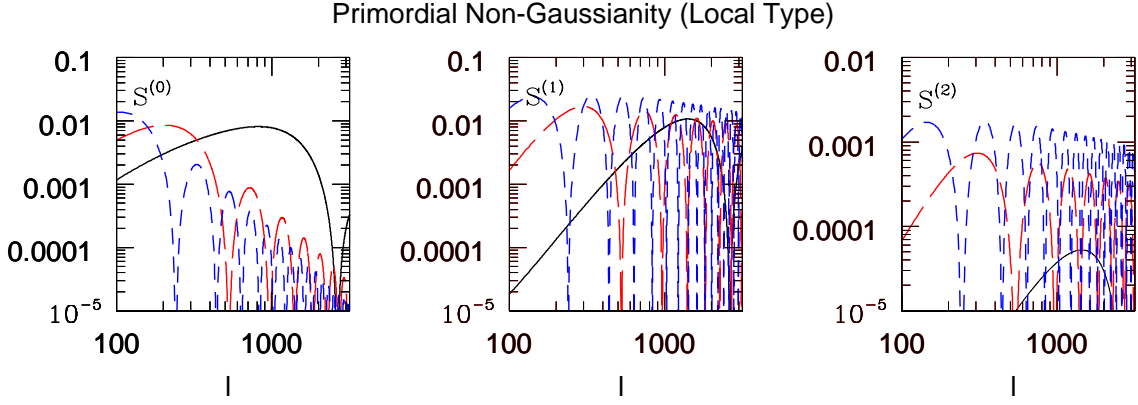}}}
\caption{The skew-spectra $S^{(0)}_l, S^{(1)}_l$ and $S^{(2)}_l$ are plotted for source redshift
$z_s=1.0$ as a function of wave number $l$. The skew-spectra correspond to the primordial
bispectrum of local type Eq.(\ref{eq:f_loc}). The normlaisation coefficient is set to unity $f^{loc}_{NL}=1$.
The underlying cosmology is that of WMAP7.
A tophat window is assumed. Various curves correspond to different smoothing angular scales
as indicated. The resolution is fixed at $l_{max}=2000$. Three angular scales that we plot correspond to
$\theta_s=5'$ (solid lines), $\theta_s=25'$ (long-dashed lines) and $\theta_s=55'$ (short-dashed lines) respectively.}
\end{center}
\end{figure}

\begin{figure}
\begin{center}
{\epsfxsize=16. cm \epsfysize=5.6 cm {\epsfbox[21 528 590 715]{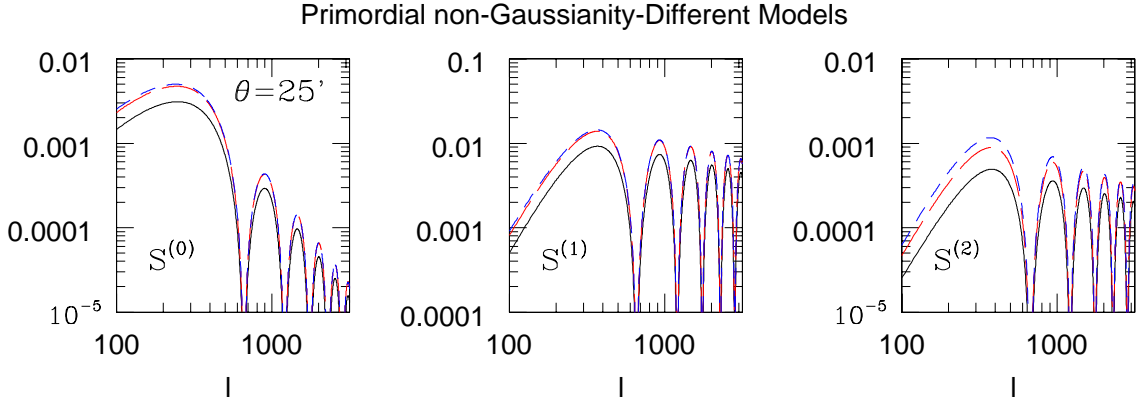}}}
\caption{The skewspectra $S^{(0)}_l, S^{(1)}_l$ and $S^{(2)}_l$ are plotted. The smoothing angular scale is $\theta_s=25'$.
The three different models that are depicted are local (solid lines), equilateral (long dashed lines)
and folded (short dashed lines) models. These models are described by Eq.(\ref{eq:f_loc}), Eq.(\ref{eq:f_equi})
and Eq.(\ref{eq:f_fol}) respectively. The redshift is fixed at unity $z_s=1$. All of the non-Gaussianity parameters
describing various models are fixed at unity i.e. $f^{loc}_{NL}=1$, $f^{equi}_{NL}=1$ and $f^{fold}_{NL}=1$.  }
\end{center}
\end{figure}

In the weakly non-linear regime ($\delta \le 1$), the description of
gravitational clustering can be described by perturbation theory
\cite{Bernardreview02}. However, the perturbative treatment breaks
down when density contrast at a given length scale becomes nonlinear
($\delta \ge 1$) which significantly increases the growth of
clustering. Perturbative studies of gravitational clustering have
attracted a lot of attention. Starting with \cite{Peebles}, there
have been many attempts to reproduce the clustering of a
self-gravitating fluid in a cosmological setting is typically
tackled by brute force using N-body simulations
\cite{Bernardreview02}. Expanding the density contrast in a Fourier
series, and assuming the density contrast is less than unity, for
the pertubative series to be convergent, we get

\be \delta({\bf k}) = \delta^{(1)}({\bf k}) + \delta^{(2)}({\bf k})
+ \delta^{(3)}({\bf k}) + \dots; \quad \delta^{(2)}(k) = \int {
d^3k_1 \over 2\pi} \int { d^3 k_2 \over 2\pi} \delta_D({\bf k_1 +
k_2 -k }) F_2(k_1,k_2) \delta^{(1)}({\bf k}_1) \delta^{(1)}({\bf
k}_2). \ee

\n The linearized solution for the density field is
$\delta^{(1)}({\bf k})$; higher-order terms yield corrections to
this linear solution. Using a fluid approach known to be valid at
large scales (and before shell crossing) one can write the second
order correction to the linearized density field using the kernel
$F_2({\bf k_1},{\bf k_2})$. Newtonian gravity coupled to the Euler
and continuity equation employed to solve a system of non-linear
coupled integro-differential equation reproduces the kernels
$F_2(k_1,k_2)$ $F_3(k_1,k_2,k_3)$ when solved perturbatively order
by order. The expression for the matter bispectrum can be written in
terms of an effective fitting formula that can interpolate between
quasilinear regime and the highly nonlinear regime:

\beqa
&& B_{\delta}(\bk_1,\bk_2,\bk_3) = 2 F_2({\bf k_1}, {\bf k_2}) P_{lin}^{\delta}(\bk_1)P_{lin}^{\delta}(\bk_2) + {\rm cyc.perm.}; \nn\\
&& F_2({\bf k_1}, {\bf k_2}) ={5 \over 7}a(n_{e},k)a(n_{e},k)+
 \left ( { {\bf k}_1 \cdot {\bf k}_2 \over 2 k_2^2}   +{ {\bf k}_1 \cdot {\bf k}_2 \over 2 k_1^2} \right ) b(n_{e},k)b(n_{e},k)
+ {2 \over 7} \left ( { {\bf k}_1 \cdot {\bf k}_2 \over k_1 k_2} \right )^2 c(n_{e},k)c(n_{e},k)
\label{eq:bispec_pert}
\eeqa

\n
The coefficients $a({n_{e}},k),b({n_{e}},k)$ and $c({n_{e}},k)$ are defined as follows:

\be
a(n_{e},k) = { 1 + \sigma_8^{-0.2}(z) \sqrt { (q/4)^{n_{e}+3.5}} \over 1 +  (q/4)^{n_{e}+3.5}}; \quad
b(n_{e},k) = {1 + 0.4(n_e+3)q^{n_{e}+3} \over 1 + q^{n_{eff}+3} }; \quad
c(n_{e},k) = { (2q)^{n_{e}+3}\over 1 + (2q)^{n_{e}+3.5} } \left \{ { 1 + \left ( 4.5 \over 1.5 + (n_{e}+3)^4 \right )} \right \}
\ee

Here $n_{e}$ is the effective spectral slope associated with the
linear power spectra $n_{e} = d \ln P_\ad / d\ln k $, q is the ratio
of a given length scale to the non-linear length scale $q=k/k_{nl}$,
where ${k^3/2\pi^2}D^2(z)P\ad(k_{nl}) = 1$ and $Q_3(n_{e})  = {(4 -
2^{n_{e}})/ (1 + 2^{n_{e}})} $. Similarly
$\sigma_8(z)=D(z)\sigma_8$. At length scales where $q \ll 1$ which
means the relevant length scales are well within the quasilinear
regime $a=b=c=1$ and we recover the tree-level perturbative results.
In the regime when $q \gg 1$ and the length scales we are
considering are well within the nonlinear scale we recover $a=
\sigma_r^{-0.2}(z) \sqrt {0.7 Q_3(n_{e})}$ with $b=c=0.$ In this
limit the bispectrum becomes independent of configuration and we
recover the hierarchical form of bispectrum discussed before.
However is there are weak violations of hierarchical ansatz in the
highly nonlinear regime is still not clear and can only be
determined higher resolution N-body simulations when they are
available. Similar fitting functions for dark energy dominated
universe calibrated against simulations are also available and at
least in the quasilinear regime most of the difference comes from
the linear growth factor \citep{Ma99}. The analytical modeling of
the matter bispectrum presented here is equivalent to the  so-called
Halo Model predictions presented above.

We have used this model to construct the analytical predictions for
various skewness parameters and the corresponding skew-spectra. The
results are plotted in  Figures \ref{fig:moments}, \ref{fig:snz},
\ref{fig:Sl} and \ref{fig:Sl_red}. In Figure \ref{fig:moments} we
have plotted the three skewness parameters $S^{(0)},S^{(1)}$ and
$S^{(2)}$ as a function of the smoothing angular scales $\theta_s$
as defined in Eq.(\ref{skewness_harmonic_space}). In
Figure-\ref{fig:snz} we change the source redshift to compare
predictions. In total we compare three different redshifts $z_s=0.5,
1.0$ and $z_s=1.5$ respectively. We use top hat filters with
different  angular smoothing scales. The skew-spectra, defined in
Eq.(\ref{skew_spectra}), are integrated measures and their value at
a specific harmonic depends on modelling of the bispectrum for the
entire range of harmonics being considered. The skew spectra are
plotted as functions of harmonic $l$ in  Figure \ref{fig:Sl} and
Figure \ref{fig:Sl_red} respectively. In Figure \ref{fig:Sl} we
consider the redshift $z_s=1.0$ and in Figure \ref{fig:Sl_red}
results for two different redshifts, $z_s=0.5$ and $z_s=1.5$, are
compared for a given angular smoothing. The oscillatory behaviour
seen in these plots is due to our choice of fiter function i.e. top
hat window.

\subsection{Primordial non-Gaussianity: bispectrum}

\begin{figure}
\begin{center}
{\epsfxsize=16. cm \epsfysize=5.6 cm {\epsfbox[31 528 590 715]{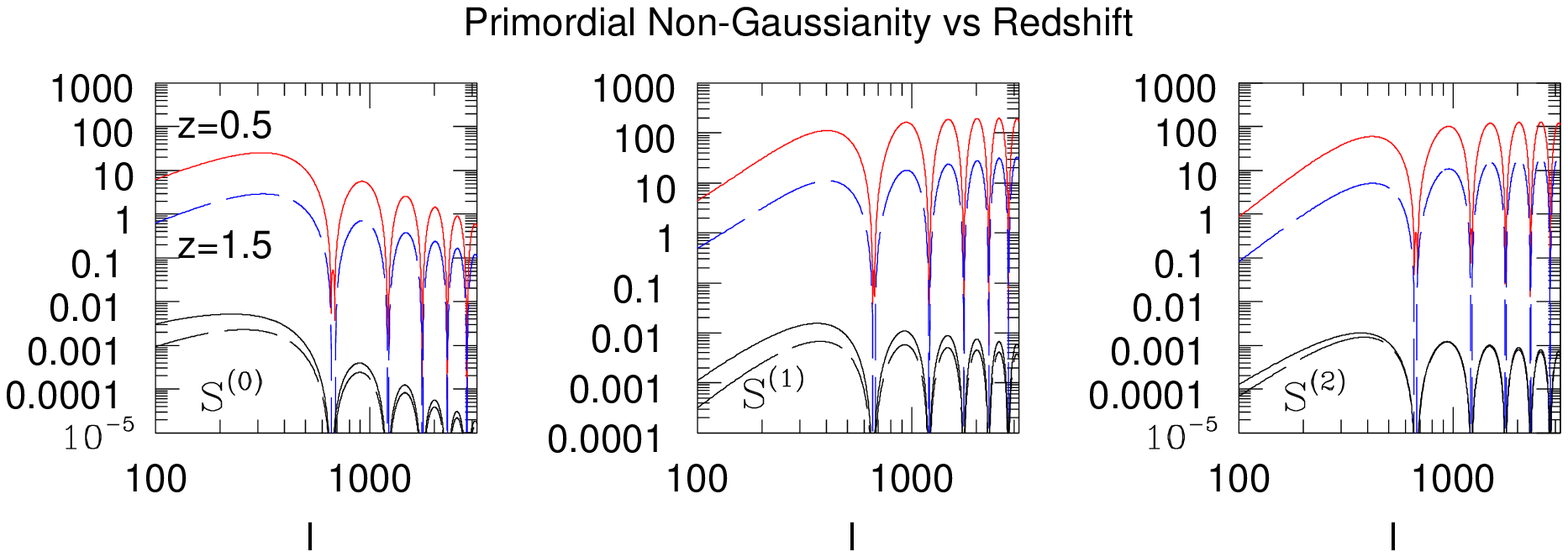}}}
\caption{The skew-spectra $S^{(0)}_l$, $S^{(1)}_l$ and $S^{(2)}_l$ are plotted for two different source redshifts
$z_s=1.5$ (long-dashed lines) and $z_s=0.5$ (solid-lines) as a function of the wave number $l$. The lower pair of curves in each panel correspond
to primordial bispectrum of local type. The gravity induced skew-spectra that dominate all scales at
each redshifts are also shown in comparison (upper pairs of curves).}
\end{center}
\end{figure}

A  recent (controversial) claim of a detection of non-Gaussianity
\citep{YaWa08} in the Wilkinson Microwave Anisotropy Probe 5-year
(WMAP5) sky maps, has boosted interest in cosmological
non-Gaussianity. Much of the interest in primordial non-Gaussianity
has focussed on a phenomenological `{\em local} $f_{NL}$'
parametrization in terms of the perturbative non-linear coupling in
the primordial curvature perturbation \citep{verdeheavens}:
\begin{equation}
\Phi(x) = \Phi_L(x) + f_{NL} ( \Phi^2_L(x) - \langle \Phi^2_L(x) \rangle ),
\label{local_real}
\end{equation}
where $\Phi_L(x)$ denotes the linear Gaussian part of the Bardeen
curvature and $f_{NL}$ is the non-linear coupling parameter. A
number of models have non-Gaussianity which can be approximated by
this form. The leading order non-Gaussianity present in this model
is at the level of the bispectrum, or in configuration space at the
three-point level. Many studies involving primordial non-Gaussianity
have used the bispectrum, motivated by the fact that it contains all
the information about $f_{NL}$ \citep{Babich}. This model has been
extensively studied
\citep{KSW,Crem03,Crem06,MedeirosContaldi06,Cabella06,SmSeZa09},
with most of these measurements providing convolved estimates of the
bispectrum. It is interesting to note here in the context of
bispectrum estimation from  CMB sky, optimized 3-point estimators
were introduced by \citet{Heav98}, and have been successively
developed \citep{KSW,Crem06,Crem07b,SmZaDo00,SmZa06} to the point
where an estimator for $f_{NL}$ which saturates the Cramer-Rao bound
exists for partial sky coverage and inhomogeneous noise
\citep{SmSeZa09}. Approximate forms also exist for {\em equilateral}
non-Gaussianity, which may arise in models with non-minimal
Lagrangian with higher-derivative terms \citep{Chen06,Chen07}. In
these models, the largest signal comes from spherical harmonic modes
with $\ell_1\simeq \ell_2 \simeq \ell_3$, whereas for the local
model, the signal is highest when one $\ell$ is much smaller than
the other two -- the so-called {\em squeezed} configuration.

In the Fourier space the
primordial bispectrum of local type defined in Eq.(\ref{local_real})
takes the following form:

\be B_\delta^{loc}(\bk_1 \bk_1,\bk_3) = 2f_{NL}^{\rm loc}\left [
P^{\Phi}_{\rm lin}(k_1) P^{\Phi}_{\rm lin}(k_2) + {\rm cyc. perm.}
\right ]. \ee

The primordial potential power spectrum in standard inflationary
models takes a power law form   $P^{\Phi}(k) \propto k^{n-4}$. In
the linear regime, the primordial bispectrum for the density field
$B_{\delta}^{\rm prim}$ in case of local model evolves according to
the following expression (see \cite{HKM06} for a detailed derivation
and discussion):

\beqa && B_\delta^{\rm loc}(\bk_1 \bk_1,\bk_3; z) = {2f_{NL}^{\rm
loc} \over D(z)}\left [ {{\cal M}({k_3}) \over {\cal M}({k_1}) {\cal
M}({k_2})} P^\delta_{\rm lin}(k_1,z)
P^\delta_{\rm lin}(k_2,z) + {\rm cyc. perm.} \right ]; \quad \\
&& \delta({\bf k},z) = D(z){\cal M}(k)\Phi({\bf k}, z); \quad {\cal M}(k) \equiv - {2 \over 3 H_0^2 \Omega_m} k^2 T(k).
\label{eq:f_loc}
\eeqa

\n Here $D(z)$ is the linear growth factor normalised such that
$D(z)\rightarrow {1 /(1+z)}$  and $T(k)$ is the transfer function
given by an approximate expression found in \citet{BBKS}. According
to standard inflationary predictions $P^{\Phi}(k) \propto k^{n-4}$
and the linear power spectra for the density is given by
$P^{\delta}_{lin}(k,z) = D^2(z){\cal M}(k)^2 P^{\Phi}(k)$. The
primordial bispectrum for the density can similarly be expressed in
terms of that of the primordial potential perturbations
$B_{\delta}^{prim}(k_1,k_2,k_3,z)= D^3(z){\cal M}(k_1){\cal
M}(k_2){\cal M}(k_3)B_{\Phi}^{prim}(k_1,k_2,k_3)$. The primordial
potential bispectrum for the equilateral type can be expressed as
\citep{Crem06}:

\be
B^{equi}_{\phi} = 6 f^{equi}_{NL} \left [ -(P_{\Phi}(k_1)P_{\Phi}(k_2) + cyc.perm.) -
2 (P_{\Phi}(k_1)P_{\Phi}(k_2)P_{\Phi}(k_3))^{2/3} +
( P^{1/3}_{\Phi}(k_1)P^{2/3}_{\Phi}(k_2)P_{\Phi}(k_2) + cyc.perm.) \right ]
\ee

The primordial density bispectrum for the equilateral case
$B_\delta^{equi}$ can be expressed, following the same procedure
that we followed for the local type, as:

\begin{eqnarray}
B_\delta^{equi}(\bk_1 \bk_1,\bk_3; z) = && {6f_{NL}^{equi} \over D(z)} \Big [ -\left ( {{\cal M}{(k_3)} \over {\cal M}{(k_1)} {\cal M}{(k_2)}} P^{\delta}_{lin}(k_1,z)
P^{\delta}_{lin}(k_2,z) + cyc. perm. \right )\nn \\
&&-2 \left ( {{\cal M}{(k_1)} {\cal M}{(k_2)} {\cal M}{(k_3)}} \right )^{-1/3} \{ {P^{\delta}_{lin}(k_1,z)P^{\delta}_{lin}(k_2,z)P^{\delta}_{lin}(k_3,z)} \}^{2/3} \nn \\
&& + \left ({{\cal M}{(k_1)}^{1/3} \over {\cal M}{(k_2)}^{1/3} {\cal M}{(k_3)}} \{ {{P^{\delta}_{lin}(k_1,z)P^{\delta}_{lin}(k_2,z)^2P^{\delta}_{lin}(k_3,z)^3} \}^{1/3}+
cyc. perm.}  \right ) \Big ].
\label{eq:f_equi}
\end{eqnarray}

\n However in contrast to the local model \label{eq:limber_approx1},
in the equilateral model the functional form of the expressions do
not have any connection to fundamental physics but are just fits
where the exact expressions are more complicated. The folded or
flattened model that is maximized for $k_2\approx k_3 \approx k_1/2$
is well approximated by the following form:

\be
B^{fold}_{\phi} = 6 f^{fold}_{NL} \left [ (P_{\Phi}(k_1)P_{\Phi}(k_2) + cyc.perm.) +
3 (P_{\Phi}(k_1)P_{\Phi}(k_2)P_{\Phi}(k_3))^{2/3}
-( P^{1/3}_{\Phi}(k_1)P^{2/3}_{\Phi}(k_2)P_{\Phi}(k_2) + cyc.perm.) \right ]
\ee

\n In terms of the density perturbations, we get the following
expression:

\begin{eqnarray}
B_\delta^{fol}(\bk_1 \bk_1,\bk_3; z) = && {6f_{NL}^{fol} \over D(z)} \Big [ \left ( {{\cal M}{(k_3)} \over {\cal M}{(k_1)} {\cal M}{(k_2)}} P^{\delta}_{lin}(k_1,z)
P^{\delta}_{lin}(k_2,z) + cyc. perm. \right )\nn \\
&& 3 \left ( {{\cal M}{(k_1)} {\cal M}{(k_2)} {\cal M}{(k_3)}} \right )^{-1/3} \{ {P^{\delta}_{lin}(k_1,z)P^{\delta}_{lin}(k_2,z)P^{\delta}_{lin}(k_3,z)} \}^{2/3} \nn \\
&& - \left ({{\cal M}{(k_1)}^{1/3} \over {\cal M}{(k_2)}^{1/3} {\cal M}{(k_3)}} \{ {{P^{\delta}_{lin}(k_1,z)P^{\delta}_{lin}(k_2,z)^2P^{\delta}_{lin}(k_3,z)^3} \}^{1/3}+
cyc. perm.}  \right ) \Big ].
\label{eq:f_fol}
\end{eqnarray}

\n
The folded or flattened form of bispectrum appears in canonical single field models where the initial
Bunch-Davies vacuum is modified.

The evolution of the primordial bispectrum is different to that
generated by gravitational evolution. The angular dependence for the
gravitationally-induced bispectrum is also different. On large
angular scales, which will be probed by future weak lensing surveys,
gravitational instability may not have erased the memory of
primordial non-Gaussianity, which can provide supplementary
information to results obtained from CMB observations.

\section{Estimators and their scatter}

\n As noted above, the estimators for the skew-spectra can be most
easily computed by cross-correlating maps in the harmonic domain.
These maps are constructed in real space by applying various
derivative operators. The recovered skew-spectra will depend on the
mask, if one is present, because a mask typically introduces
mode-mode coupling. The approach we adopt here to reconstruct the
unbiased power spectra in such a case is the Pseudo-$\myC_l$ method
 \citep{Hiv}. This approach depends on expressing the
observed power spectra $\myC_l$ in the presence of mask as a linear
combination of unbiased all-sky power spectra.

The three different generalized skew spectra that we have introduced
here can be thought as cross-spectra of relevant fields. We denote
these generic fields by $A$ and $B$ and will denote the generic
skew-spectra as $S_l^{[A,B]}$. The skew-spectra recovered in the
presence of masks will be represented as $\tilde S_l^{[A,B]}$ and
the unbiased estimator will be denoted  $\hat S_l^{[A,B]}$. The
skew-spectra recovered in the presence of mask $\tilde S_l^{[A,B]}$
will be biased. However to construct an unbiased estimator  $\hat
S_l^{[A,B]}$ for the skew-spectra the following procedure is
sufficient. The derivation follows the same arguments as detailed in
\cite{MF10} and will not be reproduced here.

\be
\tilde S_l^{[A,B]} = {1 \over 2l +1} \sum_{m} \tilde A_{lm} \tilde B^*_{lm}; \quad
\tilde S_l^{[A,B]} = \sum_{l'} M_{ll'} S_l^{[A,B]}; \quad M_{ll'} = {1 \over 2l + 1}\sum_{l'l''} I^2_{ll'l''} |w_{l''}|^2; \quad \left \{ A,B \right \} \in \left \{\myf,\myf^2, (\nabla \myf\cdot \nabla \myf), \nabla^2 \myf \right\}.  \\
\label{eq:alm_est}
\ee

\n The mode-mode coupling matrix $M$ is constructed from the power
spectra of the mask $w_{l''}$ and used for estimation of unbiased
skew-spectra $\hat S_{l'}^{A,B}$. Typically the mask consists of
bright stars and saturated spikes where no lensing measurements can
be performed. The results that we present here are generic. The
estimator thus constructed is an unbiased estimator. The computation
of the scatter covariance of the estimates can be computed using
analytical methods, thereby avoiding the need of expensive
Monte-Carlo simulations. The scatter or covariance of the unbiased
estimates $\la \delta \hat S_l^{A,B} \delta \hat S_{l'}^{A,B} \ra$
is related to that of the direct estimates $\la \delta \tilde
S_l^{A,B} \delta \tilde S_{l'}^{A,B} \ra$ from the masked sky by a
similarity transformation. The transformation is given by the same
mode coupling matrix $M$:

\be
\hat S_l^{[A,B]} = \sum_{l'} [M^{-1}]_{ll'} \tilde S_{l'}^{[A,B]}; \quad
\la \delta \hat S_l^{[A,B]} \delta \hat S_{l'}^{[A,B]} \ra =  \sum_{LL'}M^{-1}_{lL} \la \delta \tilde S_{L}^{[A,B]} \delta \tilde S_{L'}^{[A.B]} \ra M^{-1}_{L'l'};
\quad \langle \hat S_l^{[A,B]} \rangle = S_l^{[A,B]}. \nn \\
\label{eq:auto_cov}
\ee

\n The power-spectra associated with the MFs are linear combinations
of the skew-spectra (see Eq.(\ref{eq:v_k})). In our approach the
power spectra associated with the MFs are secondary and can be
constructed using the skew-spectra that are estimated directly from
the data.

No construction of an estimator is complete without an estimate of
its variance. The variance or the scatter in certain situations can
be computed using Monte-Carlo (MC) simulations which are
computationally expensive. In our approach, it is possible to
compute the covariance of our estimates of various $S_l$s, i.e.
${\langle \delta S_l \delta S_{l'} \rangle}$ under the same
simplifying assumptions that higher-order correlation functions can
be approximated as Gaussian. This allows us to express the error
covariance in terms of the relevant power spectra. The generic
expression can be written

\be
[\hat V_k^{(2)}]_l =  \sum_{l'} [M^{-1}]_{ll'} [\tilde V_k^{(2)}]_l; \quad \quad
\la \delta \hat V_k^{(2)} \delta \hat V_{k'}^{(2)} \ra =
\sum_{LL'}M^{-1}_{lL} \la \delta [\tilde V_{k}^{(2)}]_l \delta [\tilde V_{k'}^{(2)}]_{l'} \ra M^{-1}_{L'l'}
\ee

\n We would like to point out here that, in case of limited sky
coverage, it may not be possible to estimate the skew-spectra mode
by mode as the mode coupling matrix may become singular and a broad
binning of the specra may be required.

\beqa
 && \langle [\delta S_l^{[X,Y]}]\delta S_{l'}^{[X,Y]}] \rangle =
f^{-1}_{sky}{2 \over 2l+1} \left [ \myC_l^{[X,X]} \myC_{l'}^{[Y,Y]}
+ [S_l^{[X,Y]}]^2  \right ]\delta_{ll'}; \quad\quad \{X,Y\} \in \{\kappa, \kappa^2, \nabla \kappa(\oh) \cdot \nabla \kappa(\oh),
\nabla^2\kappa(\oh) \}.
\label{eq:error_cov}
\eeqa

Here the fraction of sky covered by the survey is denoted by $f_{\rm
sky}$. The expressions for the skew-spectra are quoted in
$S_l^{(\kappa^2,\kappa)}$,$S_l^{(\kappa^2,\nabla^2\kappa)}$ and
$S_l^{(\nabla\kappa\cdot\nabla\kappa,\nabla^2\kappa)}$  are given in
Eq.(\ref{skew_spectra}). The expressions for covariance also depend
on a set of power spectra  i.e. $S_l^{(\kappa^2,\kappa^2)}$,
$S_l^{(\nabla^2\kappa,\nabla^2\kappa)}$,
 $S_l^{(\nabla\kappa \cdot\nabla\kappa,\nabla^2\kappa)}$ and $S_l^{\kappa,\kappa}$. These are given by the following expression:

\ben
&& \myC_l^{\nabla\cdot\nabla,\nabla\cdot\nabla} =
\sum_{l'l''} (\myC^{}_{l'}+N_{l'})(\myC^{}_{l''}+N_{l''})[l_1(l_1+1)+l_2(l_2+1)-l(l+1)]^2  I^2_{ll'l''} W_{l'}W_{l''}; \\
&& \myC_{l}^{[\myf^2, \myf^2]} = \sum_{l'l''} (\myC^{}_{l'}+N_{l'}) (\myC^{}_{l''}+N_{l''}) I^2_{ll'l''}W_{l'}W_{l''}; \quad
 \myC_{l}^{[\nabla^2 \myf, \nabla^2 \myf]} = l^2(l+1)^2(\myC_l+N_l)W_l
\label{eq:def}
\een

\n Here $\myC_l$ is the ordinary "theoretical" convergence power
spectrum defined in Eq.(\ref{ps}) that includes noise i.e.
$\myC_lW_l$ is replaced with $\myC_lW_l+ N_l$ with $N_l=4\pi
\sigma_i^2/N_{gal}$. Here $\sigma_i$ is the intrinsic ellipticity
distribution of galaxies and $N_{gal}$ is the number of galaxies per
arc-minute square. Using these equations it is possible to compute
the scatter in various skew-spectra. These results can also be
extended to take into account the cross-correlation among various
skew-spectra extracted from the same data. Using a less compact
notation we can write

\ben
&& \langle \delta S_l^{[\kappa^2,\kappa]} \delta S_l^{[\kappa^2,\kappa]} \rangle = f^{-1}_{\rm sky} {1 \over 2l + 1}
\left [ \myC_l^{[\kappa^2,\kappa^2]}\myC_l^{[\kappa,\kappa]} + [S_l^{[\kappa^2,\kappa]}]^2 \right ] \\
&& \langle \delta S_l^{[\kappa^2,\nabla^2 \kappa]} \delta S_l^{[\kappa^2,\nabla^2 \kappa]} \rangle = f^{-1}_{\rm sky} {1 \over 2l + 1}
\left [ \myC_l^{[\kappa^2,\kappa^2]}\myC_l^{[\nabla\cdot\nabla,\nabla\cdot\nabla]} +
[ S_l^{[\kappa^2,\nabla^2\kappa]} ]^2 \right ] \\
&&  \langle \delta
S_l^{[\nabla\kappa\cdot\nabla\kappa,\nabla^2\kappa]} \delta
S_l^{[\nabla\kappa\cdot\nabla\kappa,\nabla^2\kappa]} \rangle =
f^{-1}_{\rm sky} {1 \over 2l + 1}\left [
\myC_l^{[\nabla^2\kappa,\nabla^2\kappa]}\myC_l^{[\nabla\cdot\nabla,\nabla\cdot\nabla]}
+ [S_l^{[\nabla\kappa\cdot\nabla\kappa,\nabla^2\kappa]}]^2 \right].
\een

\n
The cumulative signal to noise upto a given $l_{max}$ using these expression for estimators $S^{(0)}$ can now be expressed as:

\ben
\left [\left ( {S \over N } \right )^{0}_{lmax}\right ]^2 = f_{sky}\sum_{l}^{l_{max}} (2l+1)
\left \{  (S_l^{[\kappa^2,\kappa]})^2 \over \left [ \myC_l^{[\kappa^2,\kappa^2]}\myC_l^{[\kappa,\kappa]} + (S_l^{[\kappa^2,\kappa]})^2 \right ]  \right \}
\een

\n The signal-to-noise for the other two estimators $S^{(1)}$ and
$S^{(2)}$ can be defined likewise. The different skew-spectra that
we have studied here are not completely independent. Their
covariance can be analysed using the same procedure, allowing their
joint estimation from a single data set.

\ben
&& \langle \delta S_l^{[\kappa^2,\kappa]} \delta S_l^{[\kappa^2,\nabla^2\kappa]} \rangle = f^{-1}_{\rm sky} {1 \over 2l + 1}
\left [ \myC_l^{[\kappa^2,\kappa^2]}\myC_l^{[\kappa,\nabla^2\kappa]} +  S_l^{[\kappa^2,\nabla^2\kappa]}S_l^{[\kappa,\kappa^2]}\right ] \label{eq:10_cross} \\
&& \langle \delta S_l^{[\kappa^2,\kappa]} \delta S_l^{[\nabla\kappa\cdot\nabla\kappa,\nabla^2\kappa]} \rangle = f^{-1}_{\rm sky} {1 \over 2l + 1}
\left [ S_l^{[\kappa^2,\nabla^2\kappa]}\myC_l^{[\kappa,\nabla\kappa\cdot\nabla\kappa]} +  \myC_l^{(\kappa^2,\nabla\kappa\cdot\nabla\kappa)}
 \myC_l^{(\kappa,\nabla^2\kappa)}   \right ] \\
&& \langle \delta S_l^{[\kappa^2,\nabla^2\kappa]} \delta S_l^{[\nabla\kappa\cdot\nabla\kappa,\nabla^2\kappa]} \rangle =
f^{-1}_{\rm sky} {1 \over 2l + 1}
\left [  \myC_l^{(\kappa^2,\nabla\kappa\cdot\nabla\kappa)}
 \myC_l^{(\nabla^2\kappa,\nabla^2\kappa)} + S_l^{[\kappa^2,\nabla^2\kappa]}S_l^{[\nabla^2\kappa,\nabla\kappa\cdot\nabla\kappa]} \right ]
\een

\n The above results can be generalized to compute the
cross-covariance of $S_l$ from different sources of bispectrum. The
following quantities are required to compute the necessary
cross-covariances.

\ben
&& \myC_l^{[\kappa,\nabla^2\kappa]} = -l(l+1)\myC_l;  \quad
\myC_l^{[\kappa^2,\nabla\kappa\cdot\nabla\kappa]} = \sum_{l'l''} (\myC_{l'}W_{l'} + N_{l})
(\myC_{l''}W_{l''} + N_{l''}) I^2_{ll'l''} [l'(l'+1)+l''(l''+1)-l(l+1)]; \;\\
&& \myC_l^{[\kappa,\nabla\kappa\cdot\nabla\kappa]} = -\sum_{l'l''}  [l'(l'+1)+l''(l''+1)-l(l+1)]{\cal B}_{ll'l''}J_{ll'l''}W_{l'}W_{l''}
\een

We have discussed the lowest-order departure from Gaussianity in MFs
using a third order statistic, namely the bispectrum. The
next-to-leading descriptions are characterized by the trispectrum
which is a fourth order statistics. It is possible to estend the
definition of skew-spectra to the case of kurt-spectra or the power
spectrum associated with tri-spectra. The power spectra associated
with the Minkowski Functionals can be defined completely up to
fourth order using the skew- and the kurt-spectra. However, the
corrections to leading order statistics from kurt-spectra are
sub-dominant and leading order terms are consequently sufficient to
study the departure from Gaussianity. In any case it is nevertheless
straightforward to implement an estimator which which will estimate
the power-spectrum associated with the MFs from noisy data by
including both third order and fourth order statistics; this issue
 has been dealt with in detail in \citep{MF10} in the context
of CMB sky. The same results will also be applicable for weak
lensing surveys.

In addition to the three genralised skew-spectra that define the MFs
at lowest order in non-Gaussianity, it is indeed possible to
construct additional skew-spectra that work with different set of
weights. In principle arbitrary number of such skew-specta can be
constrcuted though they will not have direct links with the
morphological properties that we have focussed on, in this paper
they can still be used as a source of independent information on the
bispectrum and can be used in principle to separate sources of
non-Gaussianity, whether  primordial or gravity induced.

\begin{figure}
\begin{center}
{\epsfxsize=15.6 cm \epsfysize=5.6 cm {\epsfbox[25 522 587 715]{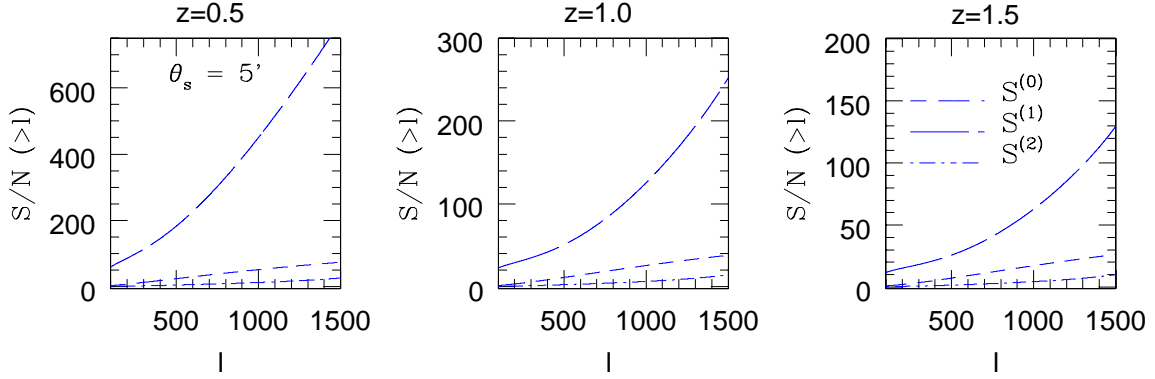}}}
\caption{The cumulative signal $S/N(>l)$ to noise associated with {\it gravity induced}
skew-spectra for $S^{(0)}_l$, $S^{(1)}_l$ and $S^{(2)}_l$ are plotted as a function of the wave number
$l$. We have assumed a full sky coverage $f_{sky}=1$. The results plotted are for $l_{max}=4000$ and
the smoothing angular scale is $\theta_s=5'$. The curves from top to bottom are $S_l^{(1)}$, $S_l^{(0)}$
and $S_l^{(2)}$ respectively. The signal-to-noise is highest for $S^{(1)}$. The non-Gaussianity
decreases with the increase in source redshift. However the power spectrum and hence the scatter
increases with redshift. This makes is easier to probe non-Gaussianity at relatively lower redshifts.}
\end{center}
\label{s2n}
\end{figure}

\section{Next to leading order corrections to the Minkowski Functionals from the trispectrum}

The skew-spectra $S^{(i)}_{l}$ and the related skewness parameters
$S^{(i)}$ completely specify the MFs at leading order. The
next-to-leading order corrections are determined by a set of four
kurtosis parameters $K^{(i)}$. These generalised kurtosis parameters
are constructed from the trispectrum using varying weights to sample
different modes. This method is very similar to construction of the
generalised skew-spectra and their associated skewness parameters
from the bispectrum described in previous sections. The kurt-spectra
are constructed by cross-correlating maps that are constructed from
original maps and  combinations of maps constructed from the
original map e.g. $\nabla \Phi(\oh)$ and $\nabla^2 \Phi(\oh)$. The
four kurtosis parameters are natural generalizations of the ordinary
kurtosis $K^{(0)}$ and can be most easily be estimated in  real
space. The normalization of these kurtosis parameters are determined
by suitable combinations of powers of parameters $\sigma_0$ and
$\sigma_1$ \citep{Mat10}.

\ben
&& K^{(0)} \equiv {1 \over \sigma_0^6}K^{(\myf^4)} = {\la \myf^4(\oh) \ra_c \over \sigma_0^6}; \quad
K^{(1)} \equiv {1 \over \sigma_0^4 \sigma_1^2 } K^{(\myf^3\nabla^2 \myf)} = {\la \myf^3(\oh) \nabla^2 \myf(\oh) \ra_c \over \sigma_0^4 \sigma_1^2}; \quad \\
&& K^{(2)} \equiv  K^{(2a)} +   K^{(2b)} \equiv {1 \over \sigma_0^2\sigma_1^4} K^{(\myf |\nabla \myf |^2 \nabla^2 \myf)} + {1 \over \sigma_0^2\sigma_1^4} K^{(|\nabla\myf|^4)}  =
{} {\la \myf |(\nabla \myf(\oh))|^2 (\nabla^2\myf(\oh)) \ra_c  \over \sigma_0^2\sigma_1^4}
+ {}{\la |(\nabla \myf)|^4 \ra_c  \over \sigma_0^2 \sigma_1^4};\\
&& K^{(3)} \equiv {1 \over 2\sigma_0^2\sigma_1^4} K^{(|\nabla\myf|^4)} = {\la |\nabla\myf(\oh)|^4 \ra_c \over 2\sigma_0^2\sigma_1^4};
\quad {\rm where} \quad |\nabla\myf(\oh)|^2 = \nabla \kappa(\oh) \cdot \nabla \kappa(\oh).
\label{kurtosis_real_space}
\een

\n Unlike the skewness parameters the kurtosis parameters get
contributions also from Gaussian (unconnected) components. The
subscript $c$ above however refers to the non-Gaussian or the
connected part of the contribution which is directly liked to the
trispectrum.

The correction to the Minkowski Functionals $\delta V^{(i)}(\nu)$ as
defined in Eq.(\ref{eq:v_k}) from the next to leading order terms
consists of both the Kurtosis parameters $K^{(i)}$ as well as the
product of two skewness parameters $S^{(i)}$ \citep{Mat10}:

\ben
&& \delta V_0^{(4)}(\nu) = {[S^{(0)}]^2 \over 72}H_5(\nu) +{K^{(0)} \over 24} H_3(\nu); \quad\\
&& \delta V_1^{(4)}(\nu) = {[S^{(0)}]^2 \over 72}H_6(\nu) +\left [ {K^{(0)}-S^{(0)}S^{(1)} \over 24} \right ] H_4(\nu) -{1 \over 12}\left [ K_1 + {3 \over 8}
[S^{(1)}]^2 \right ] H_2(\nu) - {1 \over 8} K^{(3)} \nn \\
&& \delta V_2^{(4)}(\nu) = {[S^{(0)}]^2 \over 72}H_7(\nu) + \left [ {K^{(0)} - S^{(0)}S^{(1)} \over 24} \right ] H_5(\nu) -{1 \over 6}\left[ K^{(1)} + {1 \over 2}S^{(0)}S^{(2)} \right] H_3(\nu) - {1 \over 2}\left [ K^{(2)} + {1 \over 2} S^{(1)}S^{(2)} \right ]H_1(\nu).
\label{eq:fourth_mink}
\een

The analytical modelling of four-point correlation functions is most naturally done in the harmonic domain.
They are described by the angular trispectrum $\myT^{l_1l_2}_{l_3l_4}(L)$ which is defined through the relation
$\la \kappa_{l_1m_1}\kappa_{l_2m_2}\kappa_{l_3m_3}\kappa_{l_4m_4}\ra_c = \sum_L I_{l_1l_2L}I_{l_3l_4L} \myT^{l_1l_2}_{l_3l_4}(L)$.
The trispectrum $\myT^{l_1l_2}_{l_3l_4}(l)$ is expressed in terms of the reduced trispectrum  $P^{l_1l_2}_{l_3l_4}(l)$.
Following expression was introduced by \citep{Hu00,Hu01,huOka02} and encodes all possible inherent symmetries.

\be
\myT^{l_1l_2}_{l_3l_4}(l) = P^{l_1l_2}_{l_3l_4}(l)
+ (2l+1)\left [\sum_{l'} (-1)^{l_2+l_3}\left \{ \begin{array}{ c c c }
     l_1 & l_2 & l \\
     l_4 & l_3 & l'
  \end{array} \right \} P^{l_1l_3}_{l_2l_4}(l')
+ \sum_{l'}
(-1)^{L+L'} \left \{ \begin{array}{ c c c }
     l_1 & l_2 & l \\
     l_3 & l_4 & l'
  \end{array} \right \} P^{l_1l_4}_{l_3l_2}(l') \right ].
\label{Total_Tri}
\ee

\n The matrices in curly brackets represent $6j$ symbols which are
defined using $3j$ symbols; see \cite{Ed68} for more detailed
discussions. The entities $P^{l_1l_2}_{l_3l_4}(l)$ can be further
decomposed in terms of the {\it reduced} trispectrum
$\tau^{l_1l_2}_{l_3l_4}(l)$. A specific model for the
non-Gaussianity - either primordial or gravity-induced - has a
specific prescription for the reduced trispectrum which in turn
describes the next-to-leading-order corrections to the MFs.

\be
 P^{l_1l_2}_{l_3l_4}(l) = \tau^{l_1l_2}_{l_3l_4}(l)
+ (-1)^{\Sigma_U}\tau^{l_2l_1}_{l_3l_4}(l) + (-1)^{\Sigma_L} \tau^{l_1l_2}_{l_4l_3}(l) +
(-1)^{\Sigma_L + \Sigma_U}\tau^{l_2l_1}_{l_4l_3}(l); \quad \Sigma_L = l_1+l_2+l; \quad \Sigma_U = l_3+l_4+l.
\ee

\n In addition to the original convergence trispectra $[\myT^{(0)}]$
generally used in the literature we can define a set of four
trispectra which uses different weights to samples of modes deefined
by the quadruplet of harmonic numbers ${l_i}$.

\ben
&& [\myT^{(0)}]^{l_1l_2}_{l_3l_4}(l) =  \myT^{l_1l_2}_{l_3l_4}(l); \quad
 [T^{(1)}]^{l_1l_2}_{l_3l_4}(l) = {1 \over 4} \left [{l_1(l_1+1)+l_2(l_2+1)+l_3(l_3+1)+l_4(l_4+1)}\right ] \myT^{l_1l_2}_{l_3l_4}(l); \\
&& [\myT^{(2)}]^{l_1l_2}_{l_3l_4}(l) = {1 \over 4}\left [{ l(l+1)- (l_1(l_1+1)+l_2(l_2+1))(l_3(l_3+1)+l_4(l_4+1)) }
\right ] \myT^{l_1l_2}_{l_3l_4}(l); \\
&& [\myT^{(3)}]^{l_1l_2}_{l_3l_4}(l) = {1 \over 4}\left [ {(l_1(l_1+1)+ l_2(l_2+1)-l(l+1))(l_3(l_3+1)+l_4(l_4+1)-l(l+1))}
\right ] \myT^{l_1l_2}_{l_3l_4}(l).
\label{eq:kurt_spectra1}
\een

\n The power spectrum associated with these kurtosis parameters,
i.e. the kurt-spectrum, is defined in terms of the trispectrum,
extending the previously defined skew-spectra along fairly obvious
lines. The estimation of these kurt-spectra would be done by
cross-correlating relevant fields  used to construct the related
kurtosis in real space $K^{(i)}$:

\be K^{(i)} = \sum_{l_i}\sum_{L} [\myT^{(i)}]^{l_1l_2}_{l_3l_4}(L)
I_{l_1l_2L} I_{l_3l_4L}; \quad K^{(i)}_l = \sum_{l_i}
[\myT^{(i)}]^{l_1l_2}_{l_3l_4}(l) J_{l_1l_2l} J_{l_3l_4l}; \quad
\sum_i (2l+1)K^{(i)}_l = K^{(i)}. \label{eq:kurt_spectra2} \ee The
error and  covariance associated with these kurt-spectra can be
computed using exactly the same formalism that we described in the
context of estimation of skew-spectra. The kurt-spectra being a
power spectra will contain more information compared to the kurtosis
which is a one-point estimator. Though one may be interested in
principle to extract the entire trispectrum, it may be more
realistic to use the kurt-spectra because of the likely low
signal-to-noise associated with individual harmonic modes.

To compute the kurtosis one needs a resonable model to compute the
trispectra for the convergence field. This is typically done using
the paraphernalia of the Halo Model we introduced before. The
modelling of the gravity-induced trispectrum in the Halo Model
follows the same principle as before. It involves contributions from
one-, two-, three- and four-halo contributions and the total can be
written as:

\be
T_\delta(k_1,k_2,k_3,k_4) = T^{1h}_{\delta}(k_1,k_2,k_3,k_4)+  T^{2h}_{\delta}(k_1,k_2,k_3,k_4) + T^{3h}_{\delta}(k_1,k_2,k_3,k_4)
+ T^{4h}_{\delta}(k_1,k_2,k_3,k_4)
\ee

\n
The expressions for various contributions are listed below. These can be expressed in terms of
$I^{\beta}_{\mu}(k_1,k_2,\dots, k_\mu;z)$ defined above. Notice that two-halo (2h) term has two contributions.
In the case of one represented by $T^{2h}_{31}$ there are three points in the first halo and one in the second.
Whereas $T^{2h}_{22}$ represents two points in each halo.

\beqa
&& T^{1h}_{\delta} =  I_4^0(k_1,k_2,k_3,k_4); \\
&& T^{2h}_{\delta} =T^{31} + T^{22}; \qquad
T^{2h}_{31} = P^{\delta}_{lin}(k_1)I^1_3(k_2,k_3,k_4)I^1_{1}(k_1) +  {\rm cyc.perm.}; \qquad
T^{2h}_{22} =  P^{\delta}_{lin}(k_12)I^1_2(k_1,k_2)I^1_{2}(k_3,k_4) +  {\rm cyc.perm.} \\
&& T^{3h}_{\delta} = B^{\delta}_{lin}(k_1,k_2,k_3)I^1_2(k_3,k_4)I^1_1(k_1)I^1_1(k_2) + P^{\delta}_{lin}(k_1)P^{\delta}_{lin}(k_2) I^2_2(k_3,k_4) I^1(k_1)I^1_1(k_2) +
{\rm cyc.perm.}\\
&& T^{4h}_{\delta} = I^1_1(k_1)I^1_1(k_2)I^1_1(k_3)I^1_1(k_4)T^{\delta}_{lin}(k_1,k_2,k_3,k_4) + I^2_2(k_4)I^1_1(k_1)I^1_1(k_2)I^1_1(k_3) + {\rm cyc.perm}
\eeqa

\n
Here $P^{\delta}_{lin}$ is the linear power spectrum for the density contrast $\delta$ and  $B^{\delta}_{lin}(k_1,k_2,k_3)$ and $T^{\delta}_{lin}(k_1,k_2,k_3,k_4)$ is
the tree level expression for the bispectrum in quasilinear regime. The general expression for $I^{\mu}(k_1,k_2,\dots, k_{\mu};z)$
is quoted in the expression Eq.(\ref{eq:gen_I}). Detailed derivations and discussions of these expressions can be found in e.g.
\cite{Coo_thesis,CooSeth02}. Extension of perturbative approaches can also be employed for computation of gravity induced trispectrum.
The accuracy of any analytical modeling is more difficult for the higher order multispectra and depends by and large on
more inputs from numerical simulations. The projected tripsectrum or the convergence trispectrum
${\cal T}^{l_1l_2}_{l_3l_4}(l)$ can be expressed in terms of the underlying mass trispectrum
$T^{l_1l_2}_{l_3l_4}(l)$:

\be {\cal T}^{l_1l_2}_{l_3l_4}(l) = I_{l_1l_2l}I_{l_3l_4l}
\int_0^{r_s} dr {w^4(r,r_s) \over d^6_A(r) } T^\delta \left ({l_1
\over d_A(r)},{l_2 \over d_A(r)},{ l_3 \over d_A(r)},{l_4 \over
d_A(r)} \right ) \ee  The trispectrum for primordial non-Gaussianity
exists in the literature for the local model \label{}. The results
presented here clearly are generic and can be deployed to analyze
arbitrary models. It is worth mentioning here that while modelling
of trispectrum is relevant for computation of corrections to the
leading order terms they are also important in modelling the scatter
in computation of ordinary power spectrum. Hence the errors in
$\sigma_0$ and $\sigma_1$ e.g. will involve the one-point kurtosis
parameters $K^{(i)}$ if contributions from non-Gaussianity are taken
into account. The correlation functions that represent these
kurt-spectra in real space are constructed using derivative
operators on the original convergence map and can be useful for
surveys with smaller sky coverage.

\ben
&& K^{(0)}(\oh_1,\oh_2) \equiv  \la \kappa^2(\oh_1))\kappa^2(\oh_2) \ra_c;
\quad K^{(1)}(\oh_1,\oh_2) \equiv \la \kappa^2(\oh_1) [\kappa(\oh_2)\nabla^2\kappa(\oh_2)]\ra_c; \\
&&\quad K^{(2)}(\oh_1,\oh_2) \equiv \la\nabla \kappa(\oh_1)\cdot
\nabla \kappa(\oh_1) [\kappa(\oh_2)\nabla^2\kappa(\oh_2)] \ra_c
\quad K^{(3)}(\oh_1,\oh_2) \equiv \la\nabla \kappa(\oh_1)\cdot
\nabla \kappa(\oh_1)
[\nabla\kappa(\oh_2)\cdot\nabla\kappa(\oh_2)]\ra_c. \een These
correlation functions can be computed directly in  real space
without any harmonic decomposition.

\begin{figure}
\begin{center}
{\epsfxsize=15.6 cm \epsfysize=5.6 cm {\epsfbox[25 522 587 715]{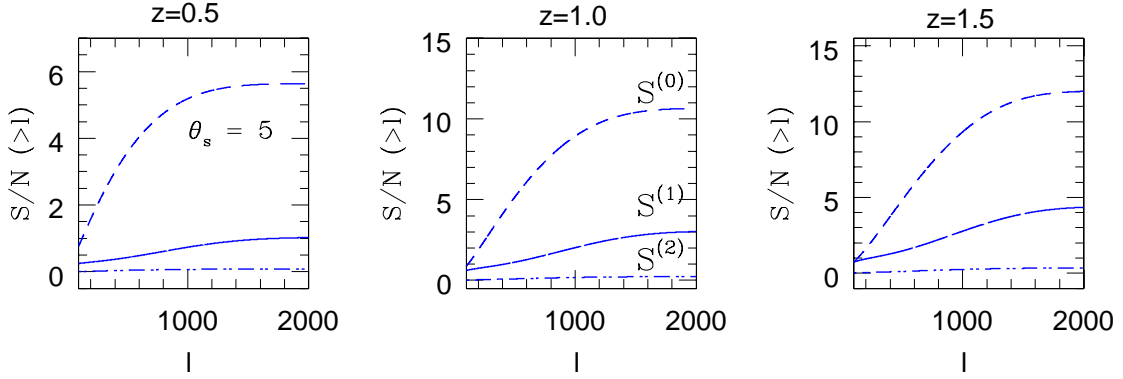}}}
\caption{Same as the previous figure but galaxy shot noise is included in the computation of scatter. }
\end{center}
\label{s2n}
\end{figure}

\section{Conclusion}

Weak lensing observations offer the potential to probe the
cosmological density distribution in an unbiased way. Since the
angular scales probed by weak lensing are sensitive to
non-Gaussianity, primarily that generated by gravitational
clustering, this technique offers us the chance to push our
understanding of the statistical properties of the cosmological
matter field far beyond current limits.

The statistical characterization of gravitational clustering is most
often performed using a hierarchy of higher order correlation
functions or their collapsed counterparts which correspond to the
moments of the convergence field $\kappa$. However, it is well known
that non-Gaussianity can also modify the morphological properties
characterized by the MFs of the relevant field $\kappa$. The MFs
therefore encode information about the non-Gaussianity and can be
used as an estimator. At leading order the MFs depend on three
generalized skewness parameters $S^0,S^1$ and $S^2$. These
parameters are one-point statistics constructed from the bispectrum
$B_{l_1l_2l_3}$  using different weights for individual modes. We
have generalized these one-point estimators to a set of power
spectra, namely $S^0_l,S^1_l$ and $S^2_l$. We studied how they can
be expressed in terms of the bispectrum $B_{l_1l_2l_3}$. In real
space these power spectra are related to the relevant correlation
function Eq.(\ref{eq:cumu_corr}). Though the correlation functions
associated with the skew-spectra are two-point statistics in terms,
of spatial order, they actually are third (lowest) order in terms of
non-Gaussianity. Hence they carry information about the bispectrum.
These statistics are in fact known as the \emph{cumulant
correlators} and the first of these statistics, $S^{(0)}$, is
already well studied in the literature. The expression for a generic
cumulant correlators of order $p+q$ is $\la
\kappa^p(\oh_1)\kappa^q(\oh_2)\ra$. It probes multispectra of order
$p+q$ and are known to be related with bias associated with over
dense objects in 3D or {\it hot-spots} in 2D \citep{Mu00}.

The skewness parameters define the leading-order terms to the MFs.
The next-to-leading-order terms are associated with the convergence
trispectrum. The convergence trispectrum in turn is expressed in
terms of trispectrum of the projected density field. The generalized
kurtosis parameters and their related power spectra can likewise be
constructed from the convergence trispectra. The corresponding
representations in the Fourier domain are named as the kurt-spectra.
We have not considered these kurt-spectra in our analysis as they
are sub-dominant but they can be taken into account using the same
formalism if required.

We have shown that the MFs can be decomposed into three different
power spectra and that these power spectra can be constructed from
an equal number of skew-spectra that carry information completely
equivalent to the original MFs at the lowest order. These power
spectra in real space will correspond to correlation functions of
fields that are constructed from products of various derivative
fields. These spatial derivative fields are in turn constructed from
the original convergence maps $\kappa(\theta_s)$. These generalized
skew-spectra are therefore related to the generalized cumulant
correlators defined in real space.  Each of these skewness
parameters can be constructed from the relevant skew-spectra.
However, the skew-spectra have the greater power in distinguishing
different sources of non-Gaussianity. This is related to the fact
that individual sources of non-Gaussiantity will lead to specific
shapes for the skew-spectra that can be tested against the observed
data. We have shown that recovery of these skew-spectra is
relatively straightforward from noisy data and in the presence of a
mask. The scatter in these statistics can be estimated under certain
simplifying approximations.

In this paper we have initiated a systematic study of these
skew-spectra in the context of weak lensing surveys. We have studied
how the skew-spectra depend on specific choices of non-linearity
that include gravity induced non-Gaussianity or  primordial
non-Gaussianity.  We have also pointed out that the departure of MFs
from Gaussianity is determined by the generalised skew-spectra which
are largley independent of cosmology but which depend primarily on
specific models of primordial non-Gaussianity. The overall
amplitudes are determined by the background cosmology as they are
determined by the power spectrum of convergence. Such a clear
distinction promises to help enormously separating the
non-gaussianity independent of cosmology.

The formalism  we have developed here for the study of
non-Gaussianity depends on the well known pseudo $C_l$ approach for
the power spectrum estimation. In this approach, the effect of any
mask and noise can be dealt with in a natural manner. This is
achieved using a matrix that encodes the mode-mode coupling. We
generalized this approach to the context of generalized skew-spectra
and showed that the error and their covariance can also be
constructed in this approach. We also performed a detailed analysis
of error characteristics. The analytical characterization of errors
means numerical costly Monte-Carlo simulations are no longer needed
and is a further strength of this approach.

It is also worth pointing out that although we have considered three
generalized skew-spectra which are related to the MFs, it is clearly
the case that infinitely many such generalized skew-spectra can
constructed with arbitrary associated weights that are \emph{not}
directly related to MFs. However these generalized skew-spectra  can
be analyzed jointly to maximize the extraction of the information
content.

One fly in the ointment is that we do not have a complete analytical
picture of gravitational clustering. However, a number of variants
of perturbative techniques which also rely on inputs from numerical
simulations are widely in use. We are also reasonably confident that
the Halo Model is  capable of capturing basic features of
gravitational instability. We have used these approximations to
construct correspondong theoretical predictions for the
skew-spectra. We study them as a function of redshift of sources as
well as the smoothing function to check how sensitive the results
are to various assumptions about the input physics.

Non-Gaussianity induced by gravity may be the primary source of
non-Gaussianity for weak lensing probes, but recent CMB studies have
also pointed to the possibility of non-zero primordial non
Gaussianity. It is well accepted that CMB studies may be the
cleanest probes to primordial non-Gaussianity. Nevertheless,
large-scale structure probes are known to reach comparable accuracy.
It is therefore interesting to see if weak lensing observations too
can be used to detect and study various models of primordial
non-Gaussianity. Motivated by the idea that the skew-spectra might
be valuable in this direction,  we have studied to what extent the
skew-spectra can provide valuable information about various models
of primordial non-Gaussianity. We specifically studied two different
models of primordial non-Gaussianity, namely the local model and the
equilateral models of non-Gaussianity,  and compared their
contributions against the gravity-induced non-Gaussianity generated
due to subsequent evolution as a function of redshift as well as
angular harmonics.

The window function which we have considered here is top hat window.
Clearly the results can be generalized to any other windows e.g. the
$M_{ap}$ or Gaussian window functions that too are often used in
various observational situations. However the use of different
window function is not expected to change the overall conclusions.

We have ignored noise in weak lensing surveys that arises from the
intrinsic distribution of galaxy ellipticities. It is expected that
noise arising from this will somewhat dilute the signatures from the
non-Gaussianity, because if increased scatter. However, for a
reasonable  number-density of galaxies, the noise power spectrum
will overtake the convergence power spectrum beyond a harmonic mode
$l$ where saturation in signal-to-noise has already been reached and
so will not likely to change the saturation value of the cumulative
signal-to-noise. This is true for all of the estimators probed as
they reach saturation for roughly the same value of $l$ as shown in
Fig $\ref{s2n}$.

Weak lensing statistics are very sensitive to the cut-off in halo
mass  used in the calculations. We have used halos in the mass range
of $10^3 M_{\sun}$ - $10^{16} M_{\sun}$. Higher-order statistics are
typically determined by the high-end tail of the density
distribution, i.e. by regions within high mass halos. Selective
choice of a specific mass range will clearly change the detailed
result and can be incorporated in our analysis. The three different
skew-spectra that we have proposed can be used to separate up to
three different components of the non-Gaussianity. Additional
skew-spectra can be constructed which can be used for a consistency
check though they may not have any direct link to the MFs. The
results presented here are also for a single source plane, e.g.
$z_s=1$, but a realistic redshift distribution of sources can easily
be incorporated in our analysis.

To summarize, we find that for $f_{NL}=1$ which specifies the
primordial non-Gaussianity the skew-spectrum is typically two orders
of magnitude lower than the gravity induced non-Gaussianity. This is
true for all three different models of primordial non-Gaussianity
that we have probed irrespective of the source redshift. This will
mean for a reasonable value of $f_{NL}$ (say e.g. $f_{NL}\approx
100$) the gravity induced non-Gaussianity and primordial
non-Gaussianity will make nearly equal contributions to the various
skew-spectra with roughly equal signal-to-noise. The scatter does
not depend on the model of non-Gaussianity and depend only on the
power spectrum. Of the three skew-spectra studied we found that the
highest signal-to-noise is achieved by the skew-spectra $S_l^{(1)}$
followed by $S^{(0)}$. For all three redshifts we have probed we
found that $S_l^{(2)}$ has the lowest signal-to-noise and may not be
detectable even with all-sky coverage.

It is worth mentioning here finally that although we have studied
the projected or 2D morphology of large-scale structure as probed by
weak lensing surveys, it is indeed possible to extend these results
to 3D weak lensing surveys. The 3D weak lensing survey generalizes
the tomographical studies to 3D using photometric redshifts. In
future many 3D weak lensing surveys will provide us with an unbiased
picture of the dark matter distribution. Statistical descriptors
will be important to quantify such 3D distribution of dark matter.
The 3D morphology of the large scale structure has also been studied
extensively studied using morphological descriptors applied to
redshift surveys; see \citep{js06} and the references therein. The
3D morphology is far richer than the 2D descriptors considered here
for the projected surveys. In 3D there are four MFs which correspond
to the surface area $V_0$, volume $V_1$, extrinsic $V_3$ and
intrinsic curvatures $V_4$ respectively. These MFs are used to
define various statistics that are linked to genus and percolation
statistics.  {\it Shape statistics} have also been introduced to
link the MFs with statistical analysis of shapes and are now widely
used for analyzing galaxy surveys and N-body simulations.

In future, use of photometric information of galaxies will allow
mapping out the dark matter distribution using weak lensing surveys.
Such 3D weak lensing surveys will provide us 3D maps of the dark
matter distribution that can be probed using morphological
descriptors. The direct link with bi- and tri-spectra based approach
developed here can be useful in studying growth of structure under
gravitational instability. These has the potential to greatly
enhance the information gained by studying projected catalogs that
we have presented here.

\section{Acknowledgements}
\label{acknow} DM acknowledges support from STFC standard grant
ST/G002231/1 at School of Physics and Astronomy at Cardiff
University where this work was completed. LVW is supported by NSERC and CIfAR.
JS acknowledge support from NSF AST-0645427.
\bibliography{paper.bbl}

\appendix

\end{document}